\documentclass[iop,revtex4]{emulateapj}
\usepackage{epsfig}
\usepackage{amsmath}
\usepackage{natbib}
\usepackage{multirow}
\usepackage{rotating}
\usepackage{graphicx}

\def\jj{$J_{\rm 125}$} 
\def\yy{$Y_{\rm 105}$} 
\def\jh{$JH_{\rm 140}$} 
\def\hh{$H_{\rm 160}$} 

\def\oo2{[$\rm O\,II$]}
\def\o3{[$\rm O\,III$]}

\def\BORG{BoRG[$z$9-10]}

\newcommand\Tstrut{\rule{0pt}{2.6ex}}       
\newcommand\Bstrut{\rule[-0.9ex]{0pt}{0pt}} 

\shorttitle{The $z\sim9-10$ BoRG {\it HST} survey}
\shortauthors{Calvi V. et al.}

\begin{document}

\title{Bright Galaxies at Hubble's Redshift Detection Frontier: 
  Preliminary Results and Design from the redshift \lowercase{{\it z}} $\sim9-10$
  B\lowercase{o}RG pure-parallel {\it HST} survey}

\author{V. Calvi\altaffilmark{1}, M. Trenti\altaffilmark{2},
  M. Stiavelli\altaffilmark{1}, P. Oesch\altaffilmark{3,4},
  L. D. Bradley\altaffilmark{1}, K. B. Schmidt\altaffilmark{5,6},
  D. Coe\altaffilmark{1}, G. Brammer\altaffilmark{1}, S.
  Bernard\altaffilmark{2}, R. J. Bouwens\altaffilmark{7},
  D. Carrasco\altaffilmark{2}, C. M. Carollo\altaffilmark{8},
  B. W. Holwerda\altaffilmark{7}, J. W.  MacKenty\altaffilmark{1},
  C. A. Mason\altaffilmark{5,9}, J. M. Shull\altaffilmark{10}, and T. Treu\altaffilmark{9}}
\date{\today}

\altaffiltext{1}{Space Telescope Science Institute, 3700 San Martin
  Drive, Baltimore, MD 21218, USA. Email: calvi@stsci.edu;
  valentinacalvi86@gmail.com} \altaffiltext{2}{School of Physics,
  University of Melbourne VIC 3010, Australia. Email:
  michele.trenti@unimelb.edu.au} \altaffiltext{3}{Yale Center for
  Astronomy and Astrophysics, Physics Department, New Haven, CT 06520,
  USA.}  \altaffiltext{4}{Department of Astronomy, Yale University,
  New Haven, CT 06520, USA.}  \altaffiltext{5}{Department of Physics,
  University of California, Santa Barbara, CA 93106-9530, USA.}
\altaffiltext{6}{Leibniz-Institut fur Astrophysik Potsdam (AIP), An
  der Sternwarte 16, 14482 Potsdam, Germany.}  \altaffiltext{7}{Leiden
  Observatory, Leiden University, NL-2300 RA Leiden, The Netherlands.}
\altaffiltext{8}{Institute of Astronomy, ETH Zurich, CH-8093 Zurich,
  Switzerland.}  \altaffiltext{9}{Department of Physics and Astronomy,
  UCLA, Los Angeles, CA, 90095-1547, USA.}  \altaffiltext{10}{CASA,
  Department of Astrophysical and Planetary Science, University of
  Colorado, 389-UCB, Boulder, CO 80309, USA.}

\begin{abstract}

We present the first results and design from the redshift $z\sim9-10$
Brightest of the Reionizing Galaxies {\it Hubble Space Telescope}
survey \BORG, aimed at searching for intrinsically luminous unlensed
galaxies during the first 700 Myr after the Big Bang.  \BORG\/ is the
continuation of a multi-year pure-parallel near-IR and optical imaging
campaign with the Wide Field Camera 3.  The ongoing survey uses five
filters, optimized for detecting the most distant objects and offering
continuous wavelength coverage from $\lambda=0.35\mu$m to
$\lambda=1.7\mu$m.  We analyze the initial $\sim130$ arcmin$^2$ of
area over 28 independent lines of sight ($\sim 25$\% of the total
planned) to search for $z>7$ galaxies using a combination of Lyman
break and photometric redshift selections. From an effective comoving
volume of $(5-25) \times 10^5$ Mpc$^3$ for magnitudes brighter than
$m_{AB}=26.5-24.0$ in the \hh-band respectively, we find five galaxy
candidates at $z\sim 8.3-10$ detected at high confidence ($S/N>8$),
including a source at $z\sim 8.4$ with $m_{AB}=24.5$ ($S/N\sim22$),
which, if confirmed, would be the brightest galaxy identified at such
early times ($z>8$). 
In addition, \BORG\/ data yield four galaxies with
$7.3\lesssim z\lesssim 8$. These new Lyman break galaxies with
$m\lesssim 26.5$ are ideal targets for follow-up observations from
ground and space based observatories to help investigate the complex
interplay between dark matter growth, galaxy assembly, and
reionization.
\end{abstract}


\section{Introduction}
\label{sec:introduction}

Early galaxies, observed when the Universe was only 500-800 Myr old,
need to be identified and studied using deep observations reaching
magnitudes $m_{AB}\gtrsim 26$ at near-infrared (near-IR)
wavelengths. Essentially, this currently limits the discovery
capabilities to observations with the {\it Hubble Space Telescope}
({\it HST}).  Still, prior to the last servicing mission (2009) {\it
  HST} imaging efficiency in the near-IR was not competitive compared
to observations in the optical both in terms of detector area and
sensitivity.  The installation of the Wide Field Camera 3 (WFC3)
on-board {\it HST} has removed this technological barrier. Whereas
only a handful of candidates at redshift $z\gtrsim7$ were known
previously
\citep{bouwens2008,bouwens2010_nicmos,bradley2008,oesch2009}, the
combination of all datasets available to search for high-$z$ galaxies
that have been acquired in the last five years now provides a sample
approaching 1000 candidates
\citep{bouwens2011,bouwens2015_10000gal,trenti2011,bradley2012,oesch2012,oesch2014,mclure2013,schenker2013,finkelstein2015,schmidt2014,bradley2014},
reaching up to $z\sim 11$ (400 Myr; \citealt{coe2013}). This
transformation has been made possible thanks to a combination of
ultradeep, small area surveys such as the UDF09 and
UDF12\footnote{Hubble Ultra Deep Field 2009, PI. G. Illingworth;
  Hubble Ultra Deep Field 2012, PI. R. Ellis} campaigns
\citep{illingworth2013,Ellis2013}, observations targeting
cluster-scale gravitational lenses, in particular
CLASH\footnote{Cluster Lensing And Supernova survey with Hubble,
  PI. M. Postman} \citep{postman2012} and the Frontier Fields
Initiative \citep{coe2015}, large area surveys over legacy fields
(CANDELS\footnote{Cosmic Assembly Near-Infrared Deep Extragalactic
  Legacy Survey, PIs: S. M. Faber, H. C. Ferguson};
\citealt{grogin2011,koekemoer2011}) and with random pointings
(BoRG\footnote{Brightest of Reionizing Galaxies, PI. M. Trenti};
\citealt{trenti2011}).

These observations are allowing a progressively more precise
characterization of the evolution of the galaxy luminosity function
(LF) in the rest-frame UV ($\lambda \sim 0.15\mu$m).  Overall,
  space-based observations indicate that the UV LF remains well
  described by a \citet{schechter1976} form, $\Phi(L) = \Phi^*
  (L/L^*)^\alpha \exp{(-L/L^*)}/L^*$, up to $z\sim 8$, similar to what
  is observed at lower redshift, but with a steepening of the
  faint-end slope $\alpha$ (e.g., \citealt{bouwens2015_10000gal}). At
  the bright end, a key open question is whether this trend continues
  into the core of the reionization epoch ($z\gtrsim 9$), when AGN
  feedback might be less effective
  (e.g., \citealt{finlator2011}). Observations to answer this question
  are however difficult because of the rarity of $L>L_*$ galaxies
  which implies the requirement of large-area surveys. In addition,
  excess variance because of large-scale structure (``cosmic
  variance'') impacts contiguous surveys (e.g., see
  \citealt{trenti2008,robertson2010}), and gravitational lensing
  magnification can alter the intrinsic shape, making a Schechter
  function look closer to a power-law
  \citep{wyithe2011,baronenugent2015,mason2015,fialkov2015}.

A second motivation for identifying the brightest galaxies during the
epoch of reionization is provided by their suitability as targets for
follow-up observations, either at infrared wavelengths with
Spitzer/IRAC to measure or set limits on galaxy ages and stellar
masses \citep{gonzalez2010,labbe2015}, or with near-IR
spectroscopy. The latter has the goals of achieving redshift
confirmation via detection of the Ly$\alpha$ emission line, and of
investigating how the Ly$\alpha$ equivalent width changes with
redshift which can be tied to the evolution of the neutral gas
fraction in the intergalactic medium (see
\citealt{treu2012}). Several groups observed bright $z\gtrsim 7$
galaxies with 8-m class telescopes
\citep{stark2010,treu2012,treu2013,finkelstein2013,schenker2013,vanzella2014_52h,pentericci2014,oesch2015,roberts-borsani2015,zitrin2015}
reaching the conclusion that detection of Ly$\alpha$ becomes
progressively more difficult as the redshift increases. However, the
latest observations hint that bright galaxies might have higher
equivalent width distributions compared to faint galaxies at $z\gtrsim
7$ \citep{oesch2015,zitrin2015,roberts-borsani2015}. This is a trend
that, if confirmed, is the opposite of what happens at $z\lesssim 6$
and might shed light on the topology of reionization and/or on the
nature of bright objects at high-$z$.

With the goals of deriving a cosmic-variance free measurement of the
number density of $L>L*$ galaxies at $z\gtrsim 8$ and identifying new
targets for follow-up observations, we present here the survey design
and preliminary results (first $\sim 25\%$ of the area) from a new
random-pointing, pure-parallel survey with {\it HST}/WFC3, optimized
for observations at the longest wavelengths accessible to {\it
  HST}. The redshift $z\sim9-10$ Brightest of the Reionizing Galaxies
(\BORG) {\it HST} survey (GO 13767, PI. M. Trenti) is a large program
aimed at searching for intrinsically bright (\hh$<27$ mag) and
unlensed galaxies during the first 700 Myr in the history of the
Universe. \BORG\/ is complementary to the UDF and Frontier Fields
datasets, which are primarily identifying galaxies with intrinsic
luminosity $L<L_*$. In addition to exploring a new parameter space at
$z>8$, \BORG\/ data also allow us to continue increasing the sample of
bright $z\sim 7-8$ galaxy candidates, overall contributing to
preparing a sample of excellent targets for follow-up observations
during the initial stages of the {\it James Webb Space Telescope}
({\it JWST}) mission.

This paper is organized as follows: in Section \ref{sec:survey} we
describe the design of the \BORG\/ survey, in Section
\ref{sec:data_reduction} we present our data reduction pipeline,
optimized for pure-parallel (undithered) observations, and evaluate
the data quality by comparison with dithered data. Section
\ref{sec:selection} introduces the selection criteria for high-$z$
candidate galaxies from multi-band photometry, with high-confidence
candidates at $z>7$ discussed in Section \ref{sec:candidates}.  The
resulting constraints on the UV LF at $z\gtrsim8$ are presented in
Section \ref{sec:LF}, with Section \ref{conclusion} summarizing and
concluding our findings. In Appendix A we include a brief discussion
of additional dropout candidates that, while satisfying the high-$z$
selection criteria, have a higher chance of being passive galaxies at
$z\sim 2$ with colors similar to dropout galaxies at $z\gtrsim 7$.

Throughout this paper we will use the AB magnitude system
\citep{oke1983} and \citet{planck2015cosmo} cosmology.

\section{Design of the survey}
\label{sec:survey}
 
\begin{figure}[!t]
\center
\includegraphics[scale=0.45]{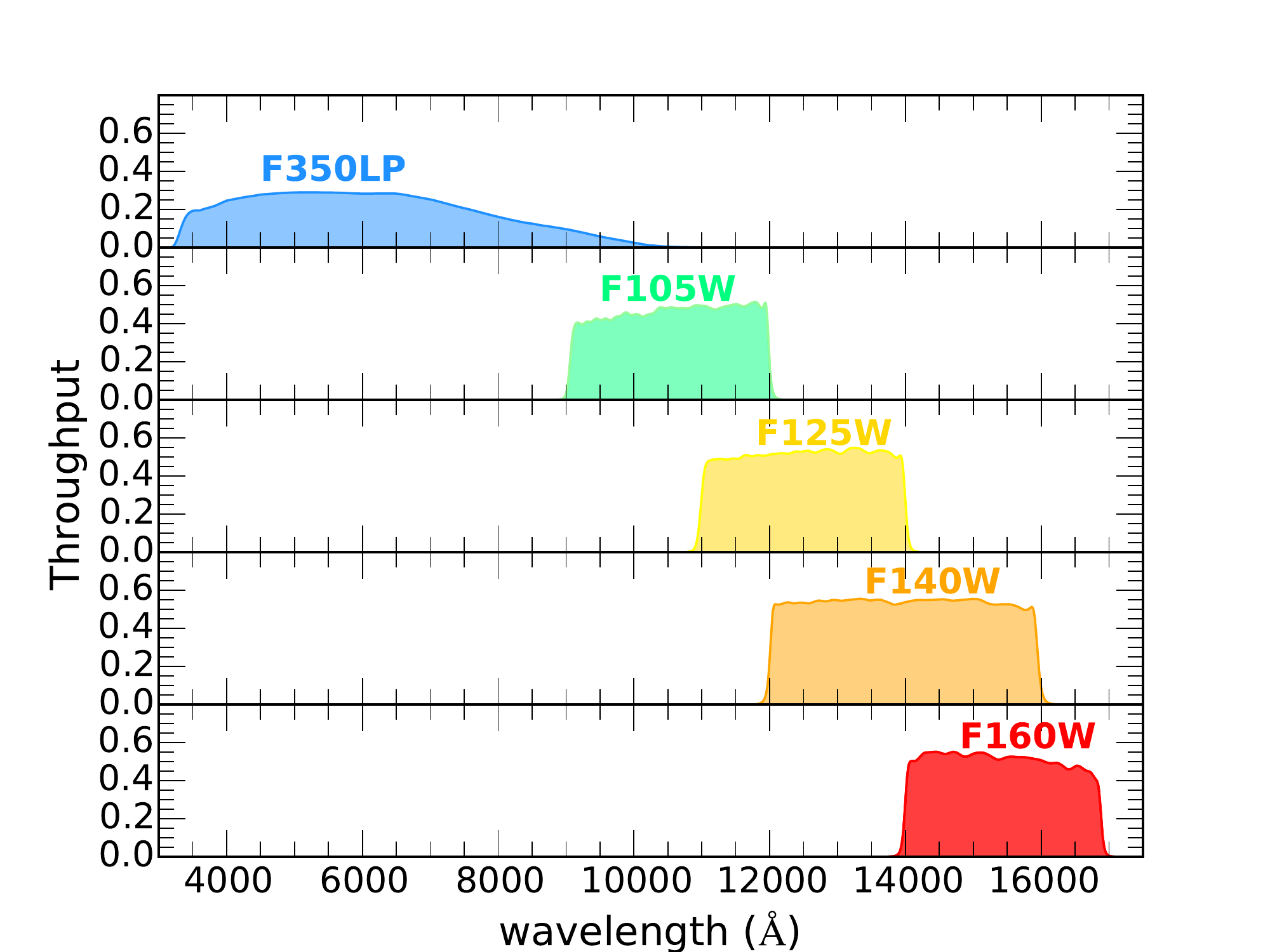}
\caption{Transmission curves of our filter set, from visible to IR:
  F350LP, F105W (\yy), F125W (\jj), F140W (\jh), and F160W (\hh) as
  labeled.  We used two complementary, non overlapping sets of IR
  filters, namely \yy-\jh\/ and \jj-\hh, for an optimal identification
  of high-$z$ galaxies through the dropout technique
  \citep{stanway2008}. }
\label{fig:wfc3_filters}
\end{figure}

The {\it HST} WFC3 \BORG\/ survey is a large (480 orbits)
pure-parallel imaging program with the nominal goal of imaging $\sim
550$ arcmin$^2$ over 120 independent lines of sight using the near-IR
filters of the Frontier Fields and HUDF12 programs: F105W (\yy), F125W
(\jj), F140W (\jh), and F160W (\hh), complemented by F350LP, a
long-pass red optical filter, achieving medium depth sensitivity
($m_{AB}\sim26.5-27.5$; $5\sigma$ point source). The main science
driver of \BORG\/ is the identification of galaxy candidates at $z>8$
from broadband colors, with a survey design optimized to constrain the
bright end of the LF at $z\sim9-10$ when the Universe was $\sim 500$
Myr old. For this design, the filter set provides continuous
wavelength coverage from $\sim 0.35\mu$m to $\sim1.7\mu$m (Figure
\ref{fig:wfc3_filters}). High-$z$ objects are selected using a
combination of the Lyman break (dropout) technique \citep{steidel1996}
and the Bayesian photometric redshift estimates (BPZ;
\citealt{benitez2000}), as discussed in Section \ref{sec:selection}.

\BORG\/ is a pure-parallel program; the WFC3 observations are carried
out while {\it Hubble} is pointed at a primary target using the Space
Telescope Imaging Spectrograph (STIS) or Cosmic Origin Spectrograph
(COS), both $\sim6$ arcmin away from WFC3 in {\it HST} focal plane. In
addition, since primary spectroscopic targets are typically in the
local Universe and/or at low redshift ($z\lesssim 3$), the volume
imaged at $z\gtrsim 6$ by WFC3 pure-parallel observations is
uncorrelated with the primary targets. WFC3 pointings in \BORG\/ have
variable exposure times, from $\sim 7000$s (3 orbits) to $\sim 19000$s
(8 orbits), with the specific duration of each opportunity determined
by the primary program.

The non-contiguous nature of a large-area survey like \BORG\/ is ideal 
for determining an unbiased measurement of the number density of
galaxies at high redshift, since these objects are strongly clustered
\citep{baronenugent2014}. In contrast, the number counts from a
contiguous large area survey are significantly affected by sample
(``cosmic'') variance, which typically introduces an additional
systematic uncertainty that is of the order of (and potentially
exceeds) the Poisson noise \citep{trenti2008}. 

The advantage of observing a large number of independent lines of
sight balances some of the challenges of pure-parallel
observations. Specifically, the depth and image quality are non uniform
across the pointings. In addition to different exposure times,
the foregrounds also vary (e.g. Galactic dust
extinction). Furthermore, pure-parallel observations are not dithered
because this would conflict with the pointing of the primary
opportunity. To minimize the impact of these limitations, we developed
a highly optimized observation design (phase II) of
\BORG\/. Specifically:
\begin{itemize}
\item We prioritized the use of primary observations that had the
  longest observing time available for parallel imaging and the least
  amount of Galactic extinction (estimated using the
  \citealt{schlafly2011} maps), although our degrees of freedom were
  limited because the pool of available opportunities (577 orbits) was
  only marginally larger than the program allocation (480 orbits).

\item To ensure high image quality (i.e to minimize spurious sources)
  and robust identification/rejection of cosmic rays, we adopted
  frequent readings (every 100s) of the near-IR detector (SPARS100
  mode), so as to have the detector well sampled. For each filter we
  also scheduled independent exposures in at least two different
  orbits. The latter choice allows us to take advantage of any
  dithering induced by small changes in the roll angle of {\it HST}
  between subsequent orbits. Observations in F350LP (optical CCD
  detector) consist of at least three independent exposures, each with
  integration time\footnote{The interval lower boundary is set to
    ensure sufficient background so that Charge-Transfer Efficiency
    (CTE) effects are not impacting the readout, while the upper
    boundary limits the number of cosmic rays present in each
    exposure.} 400s$\leq t_{exp} \leq$800s, although these might be
  scheduled in a single orbit in order to minimize the use of the WFC3
  channel select mechanism\footnote{The channel select mechanism is a
    potential non-redundant point of failure for the
    instrument. Therefore we designed the observations to use it no
    more than one time after the start of each opportunity.}.

\item We set a reference relative depth between the filters, and then
  divided the total exposure time available in each opportunity
  accordingly. Our goal is to achieve, after correcting for Galactic
  dust extinction, a near-uniform depth  in the near-IR filters,
  while F350LP observations reach 0.5 mag deeper. Specific
  opportunities may deviate from the target depths because of the need
  to satisfy the requirements discussed above on image quality and/or
  because of readout conflicts with the primary observations. A
  summary of the exposure times for each pointing analyzed in this
  paper is presented in Table \ref{tab:field_characteristics}.

\item F105W observations are scheduled in the central part of the
  orbit to minimize the impact of elevated background noise induced by
  Earth-glow \citep{brammer2014}. In addition, F160W images are taken
  last in each orbit, so as to be least impacted by any detector
  persistence from previous observations. This approach guarantees
  that any ghost source induced by persistence is brighter in the
  dropout filter compared to the detection filter, making it
  impossible to select it as a dropout candidate. This design choice
  has been used in our pure-parallel observations since Cycle 17
  \citep{trenti2011} and has demonstrated effectiveness in preventing
  the introduction of spurious dropout sources due to persistence.

\end{itemize}

The \BORG\/ design is inspired by its predecessor BoRG[$z$8]
\citep{trenti2011}, which was optimized for detection of galaxies at
$z\sim 8$ and covered about 350 arcmin$^2$ of area over 71 independent
pointings by its completion \citep{schmidt2014}. With respect to the
past survey there are two main differences: (1) the use of four IR
filters, with the addition of F140W, crucial to identify $z\sim 9-10$
galaxies, and the substitution of F098M in favor of F105W. The latter
choice is motivated by the goal of having a contiguous non-overlapping
pair of filters F105W/F140W which is optimal for selection of $z\sim
9$ galaxies \citep{stanway2008}; (2) changing F606W to F350LP to
collect more efficiently all photons at wavelengths shorter than the
CCD detector cutoff ($\sim 1\mu$m; see Figure
\ref{fig:wfc3_filters}). As discussed in Section \ref{sec:selection},
these choices are optimal for constructing clean samples of $z>8$
galaxies, but they imply an increased contamination from low-$z$
interlopers for the $z\sim 7-8$ sample (Section
\ref{sec:contamination}).

Finally, \BORG\/ represents a collection of medium-deep near IR and
optical imaging with a legacy value beyond the identification of rare
galaxies during the epoch of reionization. The high
number of independent lines of sight, distributed over a wide range of
Galactic longitudes and latitudes, has enabled the study of Galactic
structure by identifying faint M, L, and T dwarf stars
\citep{ryan2011,holwerda2014}.

\section{Data reduction}
\label{sec:data_reduction}

In this work we consider all the \BORG\/ observations acquired until
June 14, 2015, providing 42 pure-parallel opportunities. Six pairs of
opportunities cover partially overlapping regions of the sky, which we
combined to maximize the depth of the observations over the
common area, reaching up to $t_{exp}=18700$s in the deepest case. Of
these 36 independent pointings, we discarded 8 fields from the
analysis because of guide star acquisition failure\footnote{In this
  case the observations might be repeated in the future depending on
  the request by the primary observer.}, or because of excessive
stellar crowding (in case primary observations had local Universe
targets). Thus in the remainder of the paper we focus on 28
independent lines of sight suitable for searching high-$z$ galaxies,
giving a total area of $\sim 130$ arcmin$^2$. Details of each field
are provided in Table \ref{tab:field_characteristics}.

Data were downloaded from the MAST
archive\footnote{http://archive.stsci.edu/hst/search.php}, and
individual exposures processed through the standard {\tt calwf3}
pipeline to apply bias correction (UVIS only), dark subtraction, and
flat-fielding using the most up-to-date reference files. 

In addition to running {\tt calwf3}, processing of F350LP included
a correction for the Charge-Transfer Efficiency effect
(CTE\footnote{STScI CTE tools are available at
  http://www.stsci.edu/hst/wfc3/tools/cte\_tools};
\citealt{noeske2012,anderson2014}). For all filters, we performed a
customized extra step to remove residual cosmic rays and/or detector
artifacts such as unflagged hot pixels by using a Laplacian edge
filtering algorithm developed by \cite{vandokkum2001} and previously
used for BoRG[$z$8] observations \citep{bradley2012,schmidt2014}.

Despite the optimization of the orbit and filter sequence, a small
number of F105W exposures suffered from time-variable backgrounds
during the exposure \citep{brammer2014}. These backgrounds, caused by
airglow emission in the upper atmosphere, can compromise the default
up-the-ramp processing of the {\tt calwf3} pipeline and corrupt the
noise properties of the resulting calibrated images (see also
\citealt{koekemoer2013})\footnote{Note that this source of noise is
  not related to the pure-parallel nature of the observations.}.  In
these cases we remove the variable component of the background as
sampled at multiple times within the exposure and reprocess the
background-flattened sequences with {\tt calwf3} (Brammer et al., in
preparation\footnote{https://github.com/gbrammer/wfc3/blob/master/reprocess\_wfc3.py}).

To obtain the final science images in each filter we used the {\tt
  Drizzlepac} software \citep{gonzaga2012}, aligning all exposures to
a common frame and using {\tt AstroDrizzle} to construct final science
images and inverse variance maps (wht image) with a scale of 0\farcs08
pixel$^{-1}$. Since the drizzling process introduces correlated noise
regardless of the kernel used \citep{casertano2000,oesch2007}, we
derived a rescaling factor for the inverse variance maps, following the
procedure described by \citet{trenti2011}. In short, we construct a
preliminary source catalog using {\tt SExtractor}
\citep{bertinarnouts1996}, and then place empty apertures (0\farcs32
radius) over sky regions performing aperture photometry with the same
code. The errors provided by {\tt SExtractor} depend on the
variance map, defined from the inverse variance weight map as:
\begin{equation}
\frac{1}{\sqrt{\rm wht\;image}}
\end{equation}
The rms map can be rescaled by a constant factor to ensure that the
median error quoted for the photometry in an empty aperture is equal
to the variance of the sky flux measurements.  The typical rescaling
factors we applied are $1.06$ for the IR filters, and $1.33$ for
F350LP images. In addition to normalizing the variance maps, the noise
measurements done as part of this procedure allow us to quantify
the limiting magnitude of each image. The limiting magnitudes for
individual fields and filters are reported in Table
\ref{tab:field_characteristics}.

\subsection{Pure-parallel image quality}\label{sec:quality}

\begin{figure}
\centering
  \includegraphics[width=0.35\textwidth]{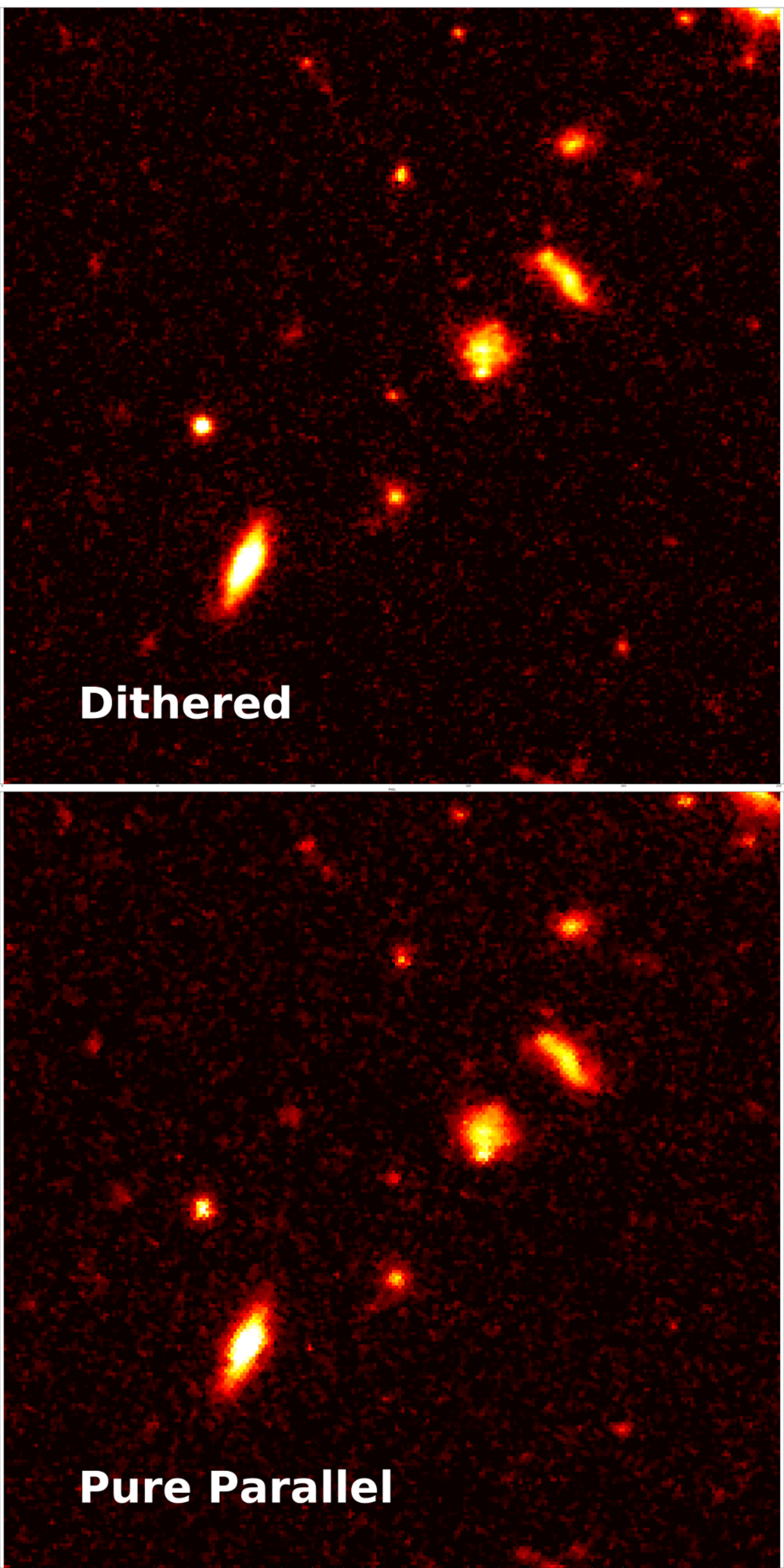}
  \includegraphics[width=0.45\textwidth]{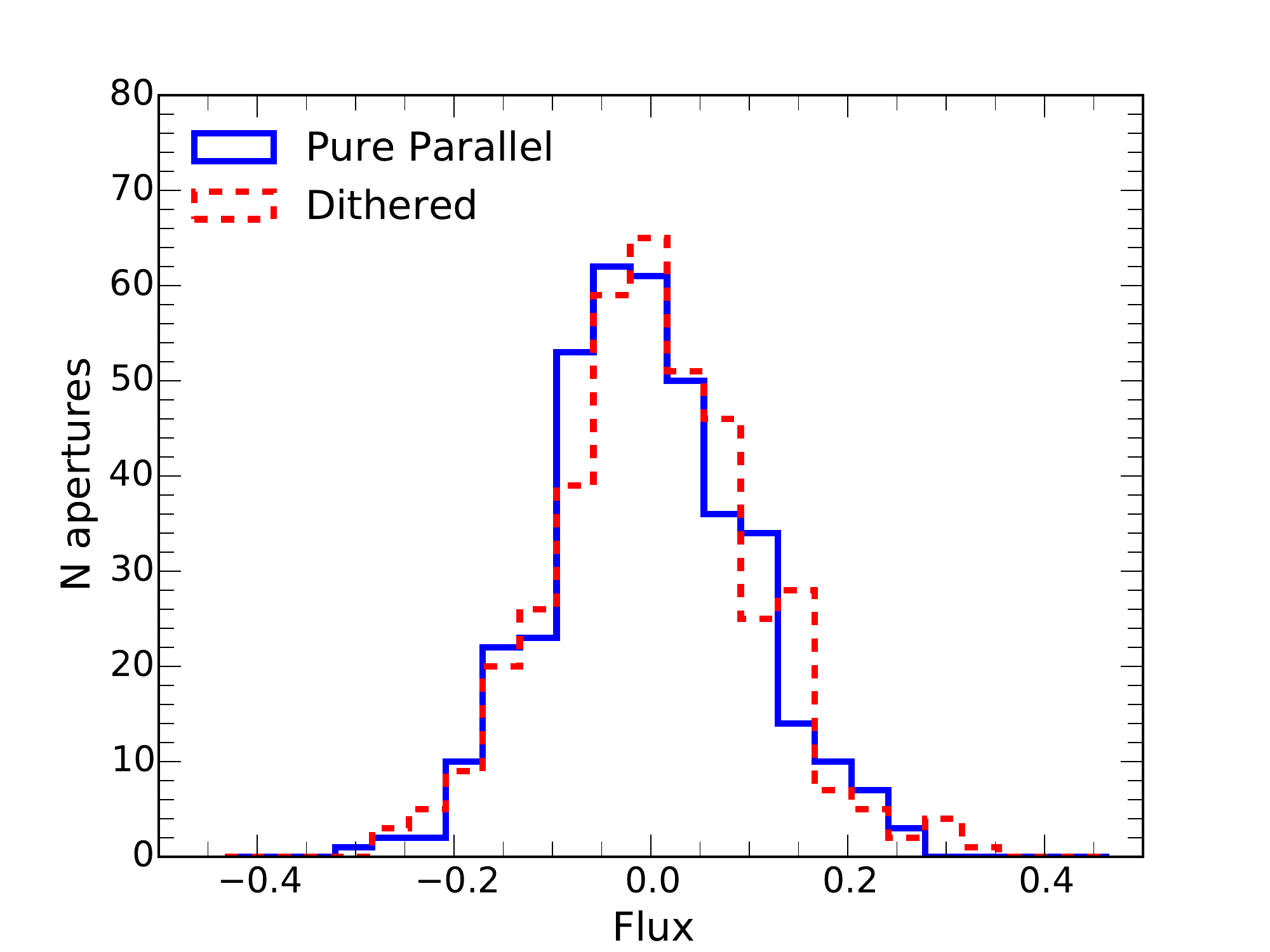}
\caption{Top panel: F125W data comparison between pure-parallel
  (bottom figure, exposure time 2500s, GO 11700) vs. dithered (top
  figure, exposure time 2300s, GO 12905) dataset for the BoRG[$z8$]
  field borg\_1437+5043 \citep{trenti2011,trenti2012,bradley2012,schmidt2014}.
  The cutout images have a $20 \farcs 0$ side and are centered on the
  bright $z \sim 8$ galaxy borg\_1437+5043\_1137 (\jj=26 mag).  Bottom
  panel: The histograms show the noise distribution in 400 empty
  apertures (sky-subtracted) with $r = 0\farcs32 $, which
  quantitatively demonstrates the near-equivalent data quality of
  pure-parallel (solid blue line) and dithered datasets (dashed red
  line).}

\label{fig:pp_vs_dith}
\end{figure}

As discussed in Section~\ref{sec:survey}, pure-parallel imaging is not
dithered, potentially affecting the data quality of the
photometry. Thanks to a follow-up program of a Y-dropout overdensity
in BoRG[$z$8] (see \citealt{trenti2012}), we have both pure-parallel
and dithered observations of overlapping area in the same IR
filters. Previously, we combined all available data to maximize the
depth of the observations, arriving at identifying the brightest
dropouts of the region at very high S/N (object borg\_1437+5043\_1137
with \jj$=25.76\pm0.07$ mag detected at $S/N_{125}\sim20$;
\citealt{schmidt2014}). Here, we re-processed the original
pure-parallel observations in F125W ($t_{exp}=2500$ s) with our latest
pipeline, and separately we analyzed the dithered (primary GO)
follow-up observations, combining a total of $t_{exp}=2300$ s of data
and using a pixel scale of 0\farcs08 pixel$^{-1}$ with the goal of
obtaining the closest analog possible to the pure-parallel
image. Visual inspection of the two science images, which are shown in
the top panels of Figure \ref{fig:pp_vs_dith}, immediately highlights
the near-equivalence of the pure-parallel data to the dithered
ones. To quantify the photometric accuracy, we selected 400 empty sky
regions in each image and performed aperture photometry (radius
$r=0\farcs32$) obtaining the noise distribution shown in the bottom
panel of Figure \ref{fig:pp_vs_dith}. The variance $\sigma$ of the
distribution for the pure-parallel and dithered dataset are 0.10 and
0.11, respectively, corresponding to $m_{lim}=28.7$ (pure-parallel)
and $m_{lim}=28.6$ (dithered) in the \jj-band, in agreement with the
exposure time calculator estimate of a $\Delta m_{lim} = 0.06$ due to
the slight difference in exposure time. Furthermore, within
statistical uncertainty, the two distributions are equivalent
(Kolmogorov-Smirnov statistics p-value 99\%). These tests demonstrate
that the lack of dithering has essentially no impact on the photometry
from pure-parallel data, which we can consider equivalent to that of
dithered data with the same exposure time within our analysis
uncertainty of $\Delta m \lesssim 0.1$.

\subsection{Source catalog construction}\label{sec:catalogs}

To construct source catalogs we ran {\tt SExtractor} in dual-image
mode. For each field, we combined all the frames taken in F140W and
F160W with {\tt AstroDrizzle} to create a detection image and a
combined weight map, which has been normalized for correlated noise
(see Section~\ref{sec:data_reduction}). As a necessary condition for
inclusion in the catalog, we required objects to have at least nine
contiguous pixels with signal-to-noise ratio per pixel $S/N \ge 0.7$.
Subsequently, we post-process the catalog to retain only sources with
isophotal $S/N\geq 8$ in the detection image.

Photometry was performed in each filter via {\tt SExtractor} dual-mode
using the detection image to define source positions and isophotal
contours. As in BoRG[$z$8], we adopt MAG\_AUTO as the total magnitude of
each source, while the signal to noise $S/N$ is defined
as:
$$\frac{S}{N}=\frac{\rm FLUX\_ISO}{\rm FLUXERR\_ISO}$$ (see
\citealt{stiavelli2009}). Finally, colors are calculated from {\tt
  SExtractor} isophotal magnitudes (MAG\_ISO),  without applying
  PSF matching, following the established practice for the BoRG survey
  (see \citealt{trenti2012}). To account for Galactic extinction in
each field, the official magnitude
zeropoints\footnote{http://www.stsci.edu/hst/wfc3/phot\_zp\_lbn}
(Zpt$_{F350LP}=26.9435$ mag, Zpt$_{F105W}=26.2687$ mag,
Zpt$_{F125W}=26.2303$ mag, Zpt$_{F140W}=26.4524$ mag,
Zpt$_{F160W}=25.9463$ mag) have been corrected using the maps by
\cite{schlafly2011}\footnote{http://irsa.ipac.caltech.edu/applications/DUST}.

To be included in the catalogs, objects need to have segmentation maps
that are associated to pixels with non-zero weight maps in all five
filters of the survey. For example, we excluded sources that fall on
the gap between the two CCD detectors and thus lack photometry in
F350LP.

Finally, we used external persistence
maps\footnote{http://archive.stsci.edu/prepds/persist/search.php},
which are released shortly after the observations, to flag any source
in the catalog that appears to be either spurious or affected by
persistent charge. Specifically, for each image and for each filter,
we created a mask that includes all pixels in the released map with
persistence value above $0.01 \mathrm{e^-/s}$, and flagged all sources
in the catalog that include at least one persistent pixel.

\subsection{Bayesian Photometric Redshifts}
\label{sec:photo-z}

For an optimal use of the full photometric information from our five
bands survey, we ran the BPZ code by \citet{benitez2000} (see also
\citealt{coe2006}) on all detected sources.  We use spectral
  energy distribution (SED) model templates as described in
  \citet{benitez2014} (but see also \citealt{rafelski2015}). Originally
  based on PEGASE models including emission lines \citep{fioc1997},
  these SEDs are recalibrated to match the observed photometry of
  galaxies with spectroscopic redshifts from FIREWORKS
  \citep{wuyts2008}. They include five early types, two late types,
  and four starbursts.  BPZ allows for interpolation between adjacent
  templates.  These 11 templates were selected to encompass the ranges
  of metallicities, extinctions, and star formation histories derived
  from galaxy observations at low and high redshift. Because of the
degeneracy between redshift and intrinsic galaxy properties such as
age, dust content, and presence of emission lines, photometric
redshift estimates for classes of rare objects with properties similar
to galaxies at $z>6$ are affected by uncertainties that are difficult
to quantify. Therefore, rather than relying only on photo-$z$ to
identify high-$z$ objects, we opt primarily for a Lyman break
selection, as discussed below. Given the challenges and
  uncertainty associated to the definition of an informed prior on the
  relative likelihood of solutions that have dropout-like colors but
  are lower-redshift interlopers, we make the minimal assumption of
  adopting a flat prior on the redshift distribution.

\section{Selection of high-redshift galaxies}
\label{sec:selection}

\begin{figure}[!h]
\center
\includegraphics[width=0.45\textwidth]{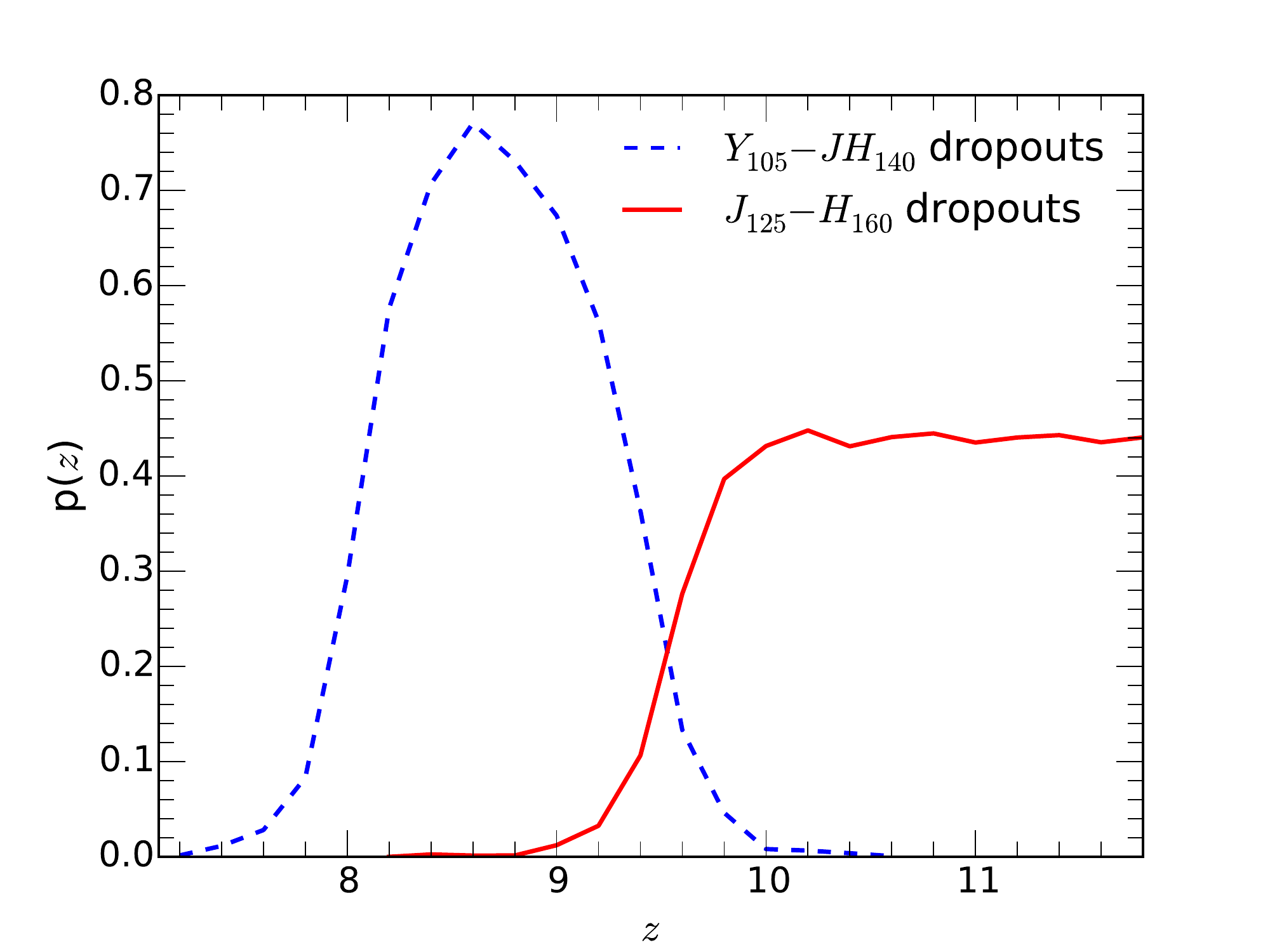}
\caption{Distribution of the probability p($z$) associated to the
  selection function as derived from our simulations (see Section
  \ref{sec:completeness}) for the field borg\_0116+1425 representative
  of a typical \BORG\/ pointing. }
\label{fig:selection_window_z9-10}
\end{figure}

From the source catalogs, we identify high-$z$ objects using the
Lyman break technique \citep{steidel1996,steidel1999}, with a set of selection
criteria similar to the ones used in legacy fields (e.g., see
\citealt{bouwens2015_10000gal}) but adapted to the specific filter set of
\BORG\/. Our general requirements are a clear detection of the source
at long wavelengths, the presence of a strong break in a pair of
adjacent, non-overlapping filters (which minimizes contamination; see
\citealt{stanway2008}), a conservative non-detection in blue bands to
reject interlopers effectively ($S/N<1.5$)\footnote{For a Gaussian
  distribution of noise, imposing a blue non-detection at $S/N<1.5$ in
  a single filter implies that we can have up to $\sim 93\%$
  completeness of the high-$z$ sample, or up to $\sim 87\%$
  completeness if the non-detection is required in two filters.}, and
a relatively flat spectrum redward of the break, again imposed to
control for contamination. In addition to $S/N\geq 8$ in the
\emph{detection image} imposed when constructing source catalogs,
these requirements translate into:

\begin{itemize}
\item For $z\sim9$ sources (\yy-\jh\/ dropouts)
$$S/N_{350}<1.5$$ 
$$S/N_{140}\ge6$$  
$$S/N_{160}\ge4$$
$$Y_{105}-JH_{140}>1.5$$
$$Y_{105}-JH_{140}>5.33\cdot(JH_{140}-H_{160})+0.7$$
$$JH_{140}-H_{160}<0.3$$ 
These selection criteria used to identify \yy-\jh\/ dropouts determine
a selection function peaked at $z=8.7$ and with a 95\% confidence
region spanning from $z=7.7$ up to $z=9.7$ (see Figure
\ref{fig:selection_window_z9-10}).

\item For $z\sim10$ sources ( \jj-\hh\/ dropouts)
$$S/N_{350}<1.5$$
$$S/N_{105}<1.5$$ 
$$S/N_{160}\ge6$$
$$J_{125}-H_{160}>1.3$$
\end{itemize}

Finally, we construct a sample of $z\sim7-8$ galaxy candidates
selected on the basis of a drop in \yy-\jj, widely used in the UDF and
CANDELS surveys, but potentially more prone to contamination in
absence of extensive multi-filter optical data because of the partial
overlap between F105W and F125W (see \citealt{stanway2008} for a
discussion of how contamination is increased by overlapping filters):
$$S/N_{350}<1.5$$ 
$$S/N_{125}\ge6$$
$$S/N_{140}\ge6$$
$$S/N_{160}\ge4$$
$$Y_{105}-J_{125}>0.45$$
$$Y_{105}-J_{125}>1.5\cdot(J_{125}-H_{160})+0.45$$
$$J_{125}-H_{160}<0.5$$

In order to avoid duplication, if a candidate satisfies more than one
selection criterion, it is assigned to the highest redshift sample.

We further refine the dropout samples by imposing a cut on the {\tt
  SExtractor} stellarity measurement, which indicates the likelihood
of having a point source, and we require CLASS\_STAR$<0.95$. Note that
this cut was not introduced primarily to reject stellar contamination,
but rather to automatically and objectively remove spurious detections
induced by hot/warm pixels that may have survived both the standard
STScI calibration and our Laplacian filtering. 

As an additional step to remove false detections from the dropouts
catalogs, VC, MT, and LB independently inspected candidates visually,
using final drizzled (and individual flt files when needed), to reject
all, and only those, sources associated to detector artifacts, hot
pixels and diffraction spikes.

Finally, to control contamination from low-$z$ interlopers in the
Lyman break samples, we ran the Bayesian photometric code BPZ
(\citealt{benitez2000}; see Section \ref{sec:photo-z} for details) and
retained in the final candidate sample only objects with photo-$z$
peaked at $z>7$.

\begin{figure*}[!t]
\center
\includegraphics[scale=0.48]{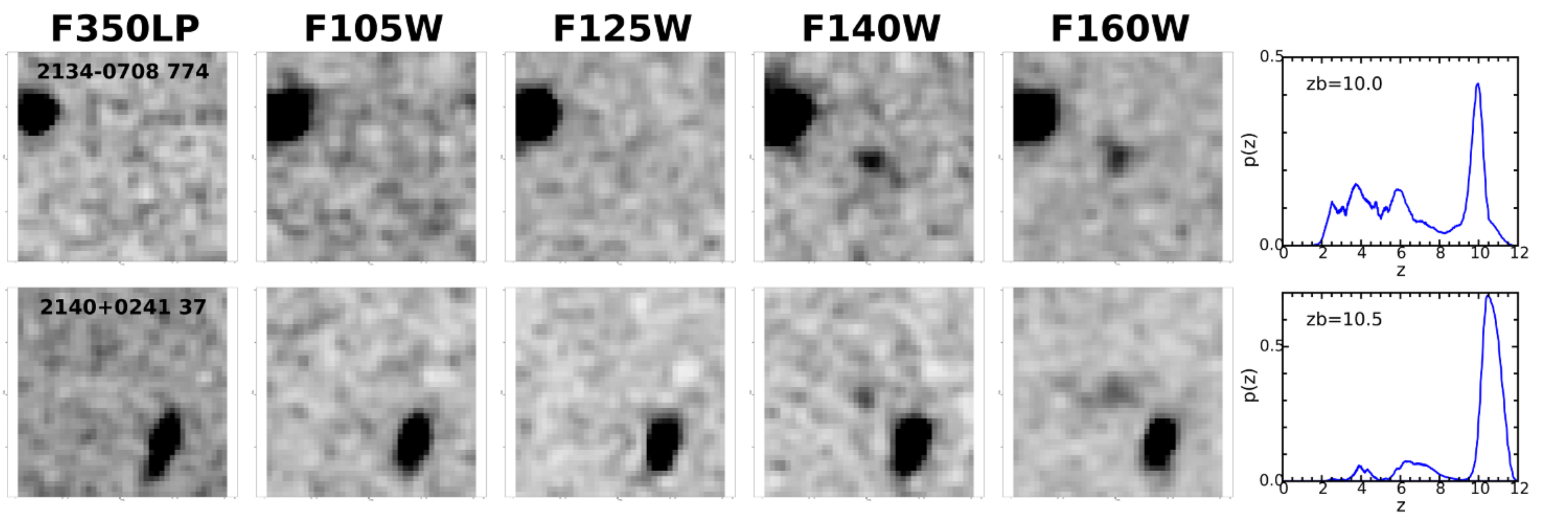}
\includegraphics[width=0.24\textwidth]{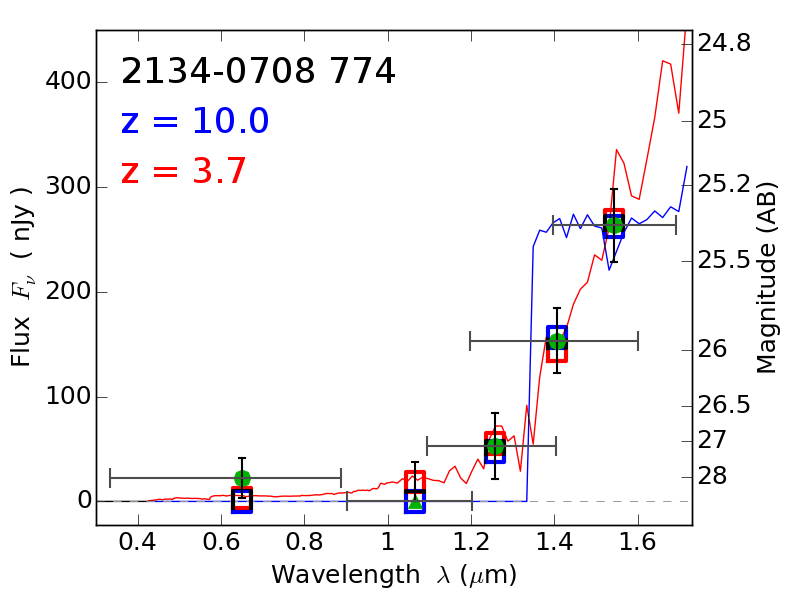}
\includegraphics[width=0.24\textwidth]{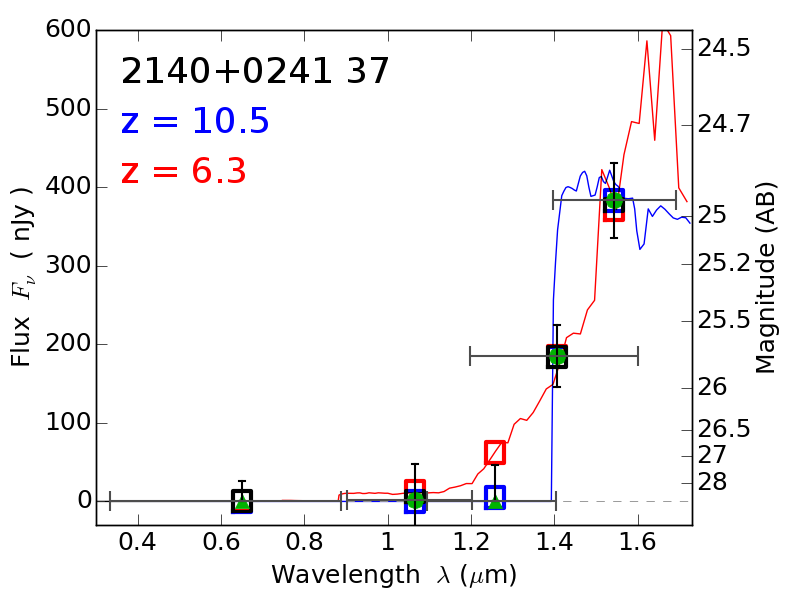}\\
\caption{Postage-stamp images of the \jj-\hh\/ dropout candidates
  listed in Table \ref{tab:z9-10}. The cutout images are $3\farcs2
  \times 3\farcs2$, each one centered on the candidate dropout galaxy.
  Right panels show the photometric redshift probability distribution
  p($z$) obtained by running BPZ using a flat prior. Bottom panels
  show the spectral energy distribution for both the the low (red) and
  high-$z$ (blue) solutions. Right axes in the SED plots show the
    total magnitudes from SExtractor MAG\_AUTO. }
\label{fig:z10}
\end{figure*}

\begin{figure*}[!t]
\center
\includegraphics[scale=0.48]{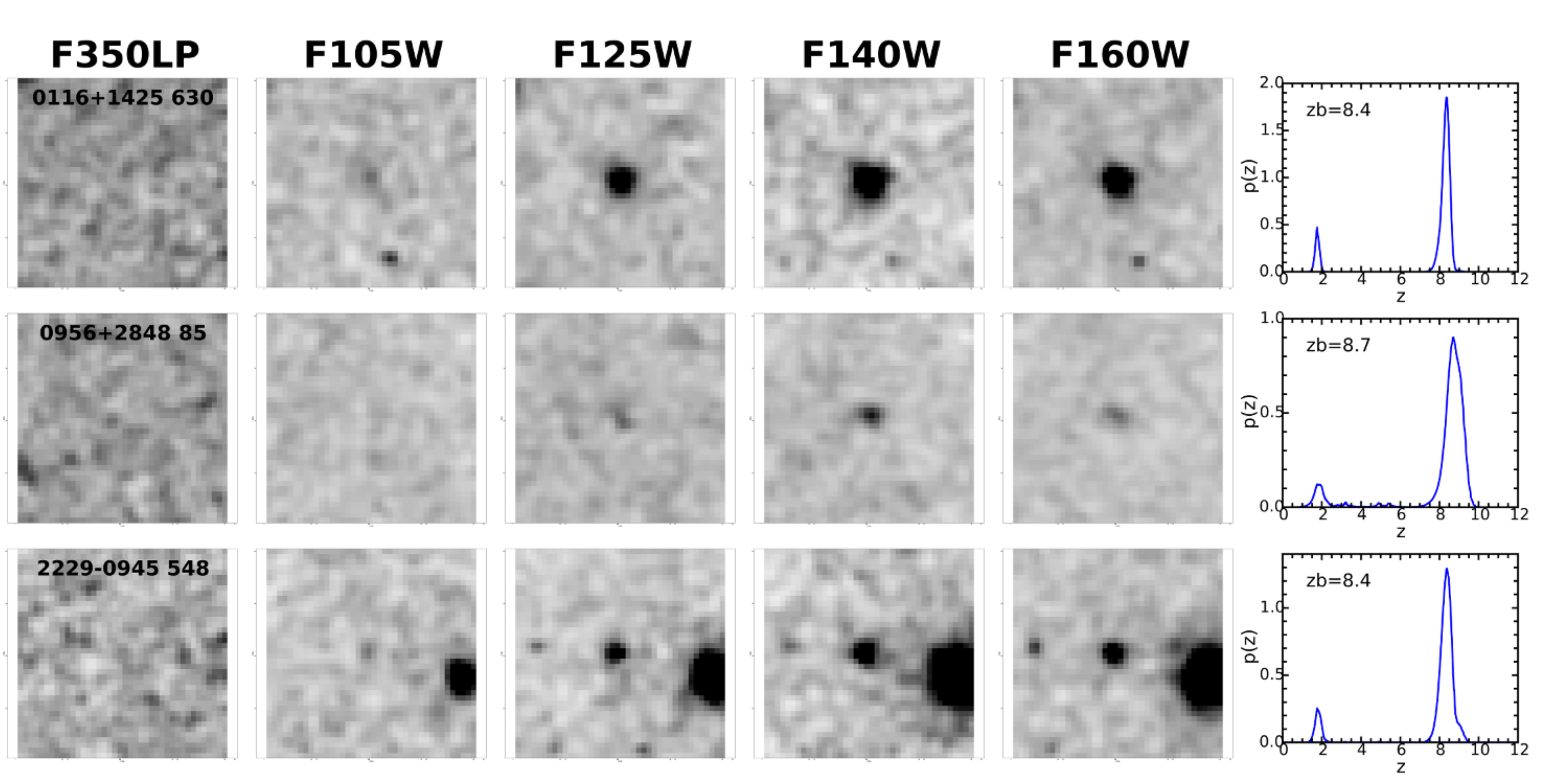}
\includegraphics[width=0.24\textwidth]{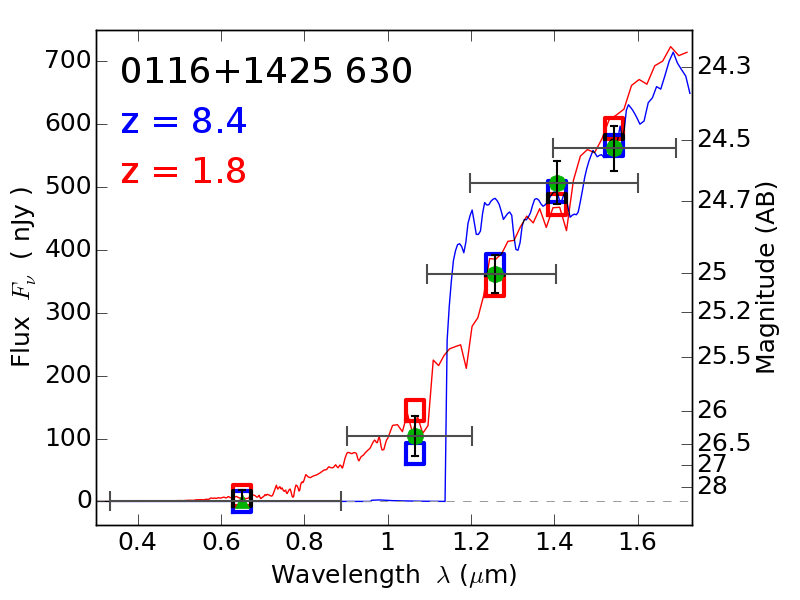}
\includegraphics[width=0.24\textwidth]{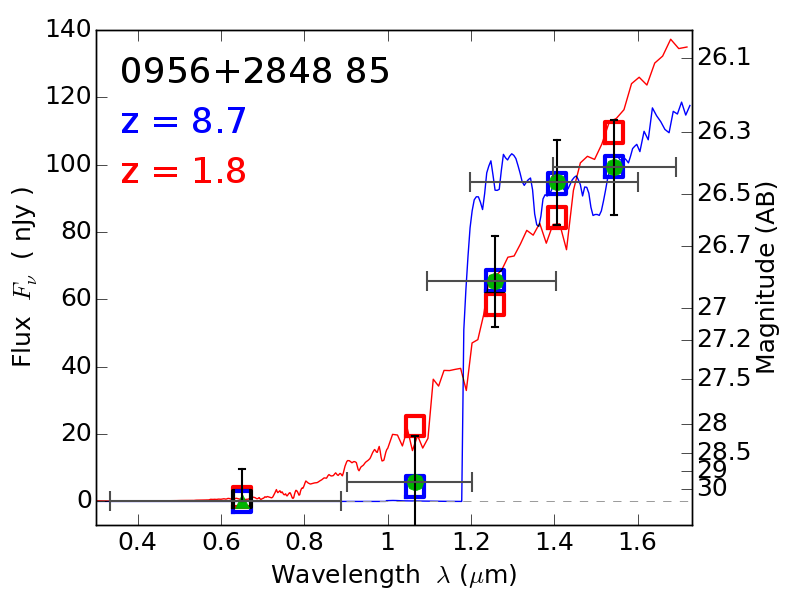}
\includegraphics[width=0.24\textwidth]{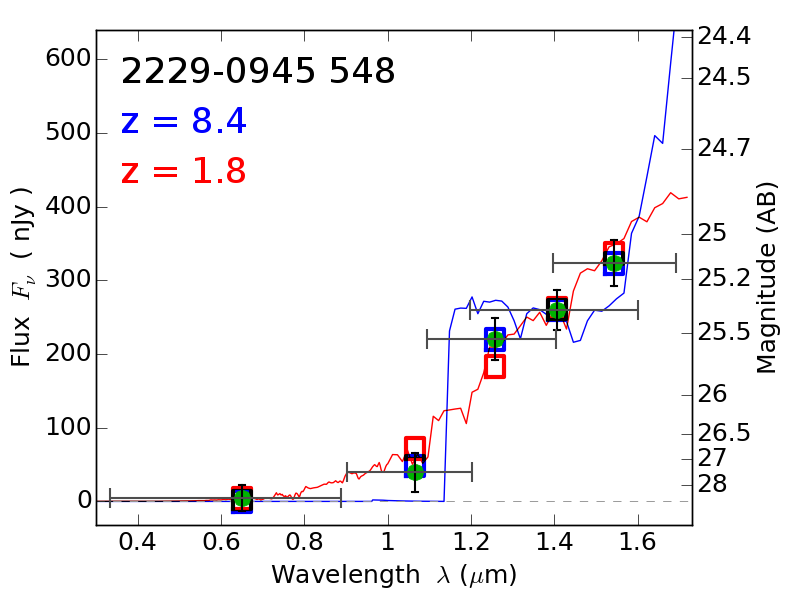}
\caption{Same as in Figure \ref{fig:z10}, but for the \yy-\jh\/ dropout candidates
  listed in Table \ref{tab:z9-10}. }
\label{fig:z9}
\end{figure*}

\begin{figure*}[!t]
\center
\includegraphics[scale=0.48]{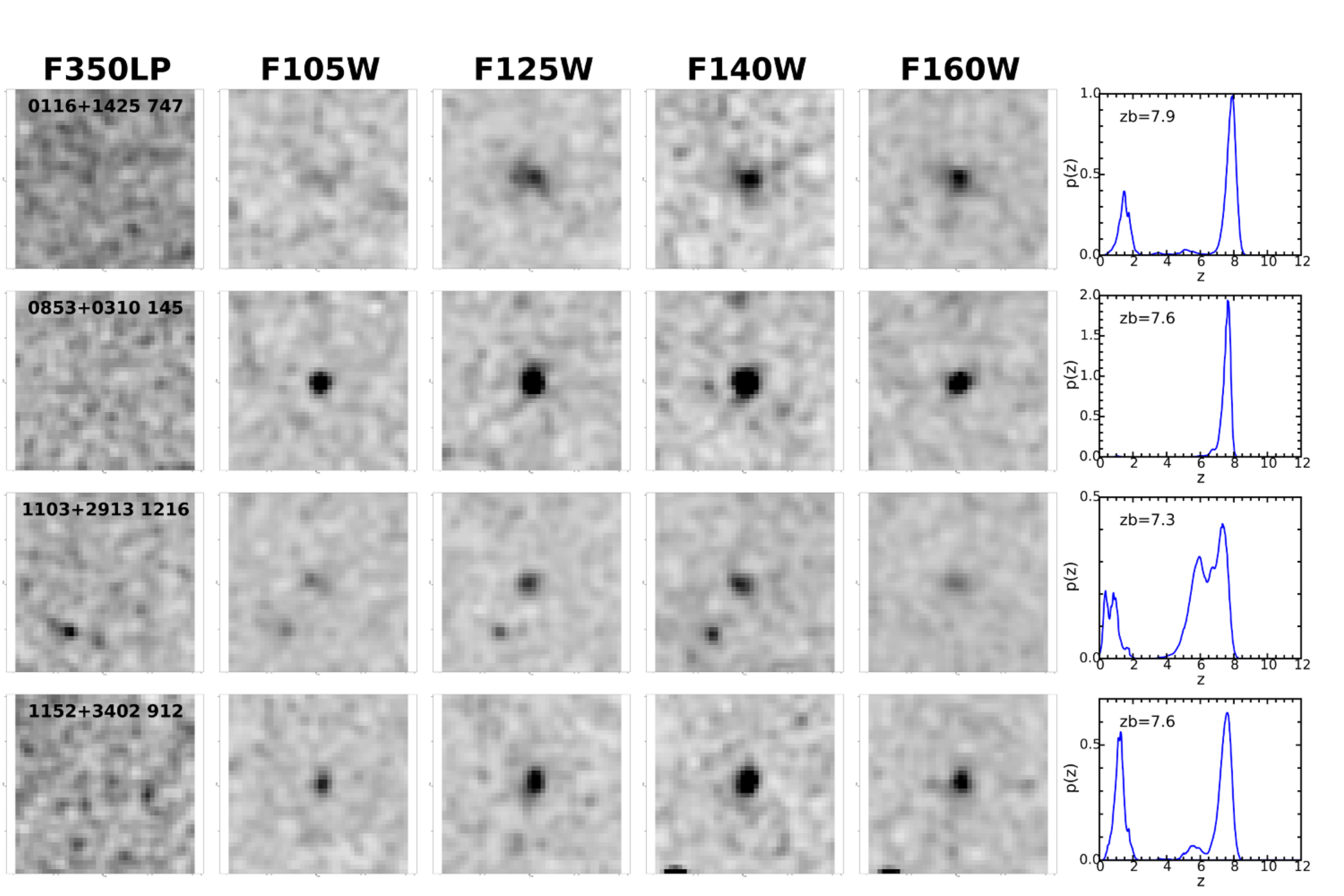}
\includegraphics[width=0.24\textwidth]{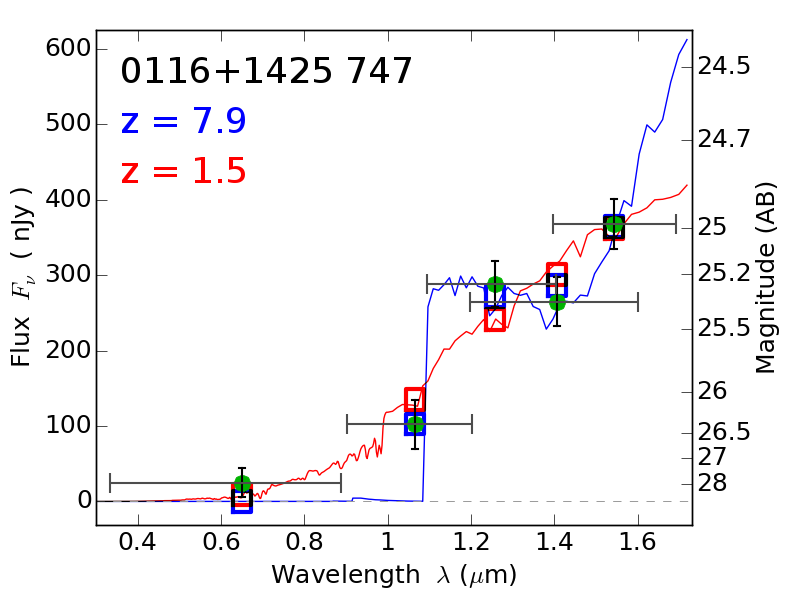}
\includegraphics[width=0.24\textwidth]{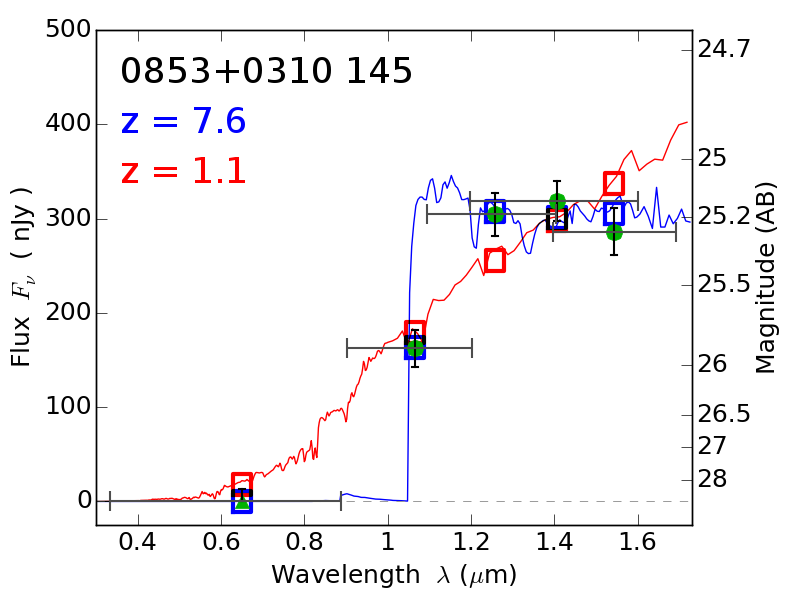}
\includegraphics[width=0.24\textwidth]{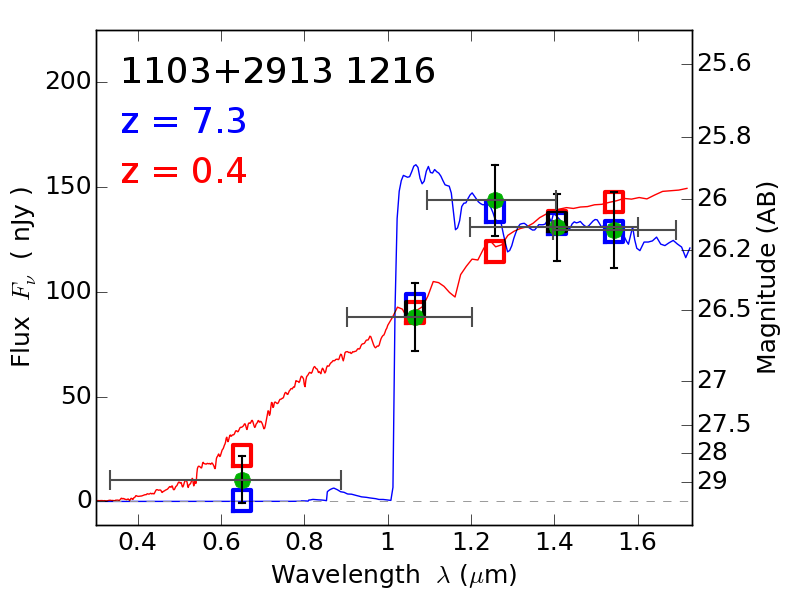}
\includegraphics[width=0.24\textwidth]{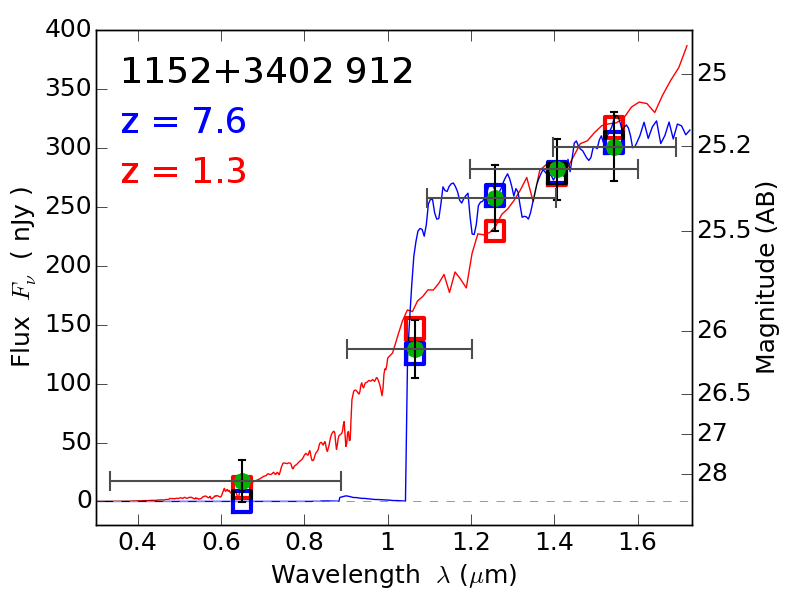}
\caption{Same as in Figure \ref{fig:z10}, but for the \yy-\jj\/ dropout
  candidates (see Table \ref{tab:z8}).}
\label{fig:z8}
\end{figure*}

\subsection{Alternative Catalogs for $z>7$ sources}
\label{sec:alternative_catalog_z7}
For the purposes of deriving LFs, we
consider as optimal our choice to construct catalogs starting from the
Lyman break selection, since it allows us to calculate the source
recovery efficiency as a function of input magnitude and redshift,
which is then used to constrain the number density of high-$z$
sources. 

However, the Lyman Break Galaxy (LBG) selection is based on a binary
decision outcome regarding inclusion of candidates in the high-$z$
source catalog, neglecting the impact of photometric uncertainties
that scatter objects in and/or out of the selection region
\citep{su2011}. Therefore, to investigate whether we are missing
objects, we employed an alternative selection by searching for sources
in our photometric catalogs that have high-$z$ solutions. Following
\citet{mclure2013}, we required non-detection in the optical
($S/N_{350}<2$) as a necessary condition, and impose stellarity
CLASS\_STAR$<0.95$, as well. Then, we selected sources that have the
peak of the redshift probability distribution function at $z>7$.

In addition, we evaluated the sensitivity of the source selection to
the construction of catalogs with a combination of F140W and
F160W. For this, we produced alternative catalogs using only F160W as
detection image. The results of this selection are summarized in
Appendix A.

\section{Results: High-\lowercase{{\it z}} Candidates}\label{sec:candidates}

\begin{figure}[!h]
\center
\includegraphics[width=0.45\textwidth]{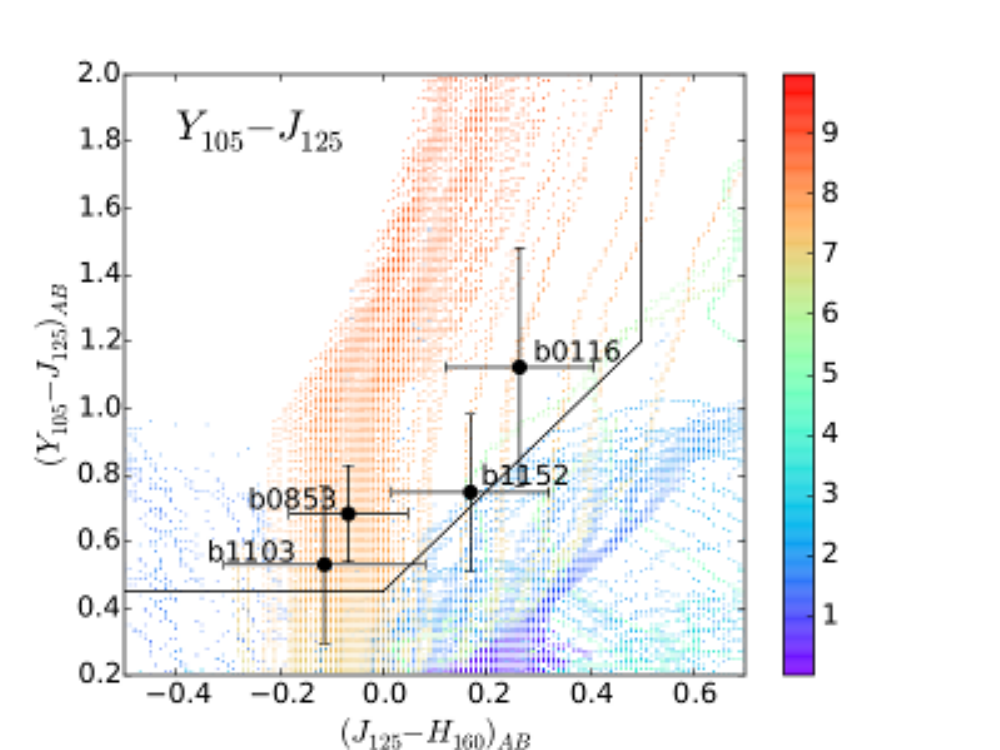}
\includegraphics[width=0.45\textwidth]{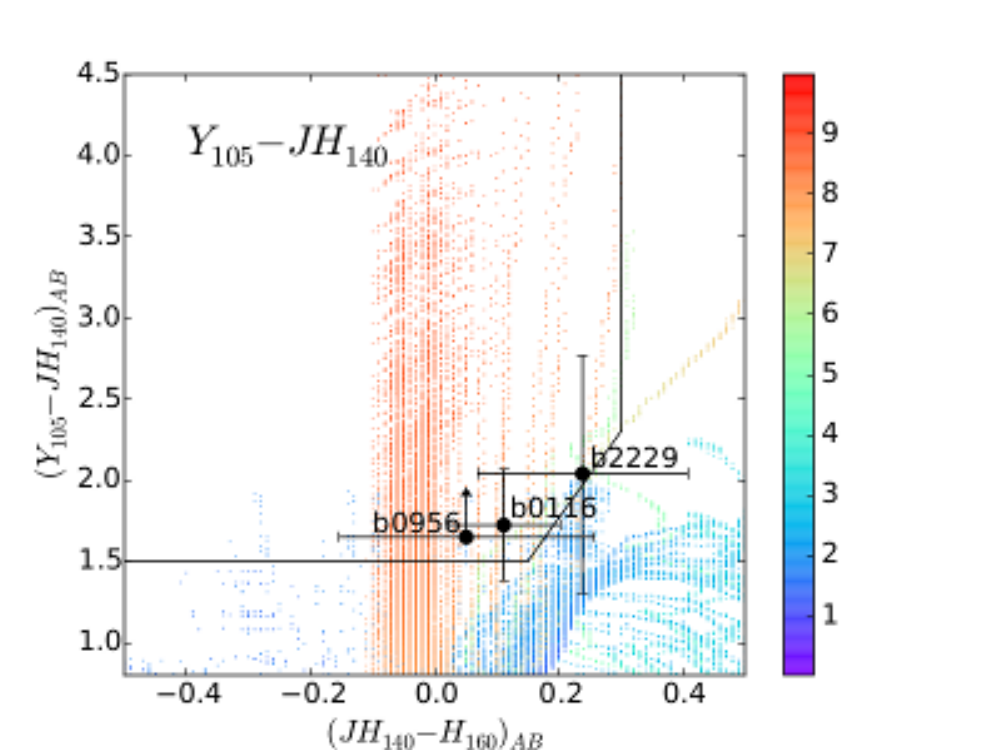}
\caption{Top panel: \yy-\jj\/ dropouts (black filled circles) in the
  \jj-\hh\/ versus \yy-\jj\/ color-color plot. The upper left region
  indicates our selection box. The colored marks show were simulated
  galaxies at different redshift (see color-bar for values)
  lie. Bottom panel: Same as in the top panel, but for the \yy-\jh\/
  dropout sources.  Colors are calculated from SExtractor
    isophotal magnitudes (MAG\_ISO).}

\label{fig:color-color_z8-9}
\end{figure}

\begin{figure}[!h]
\center
\includegraphics[width=0.45\textwidth]{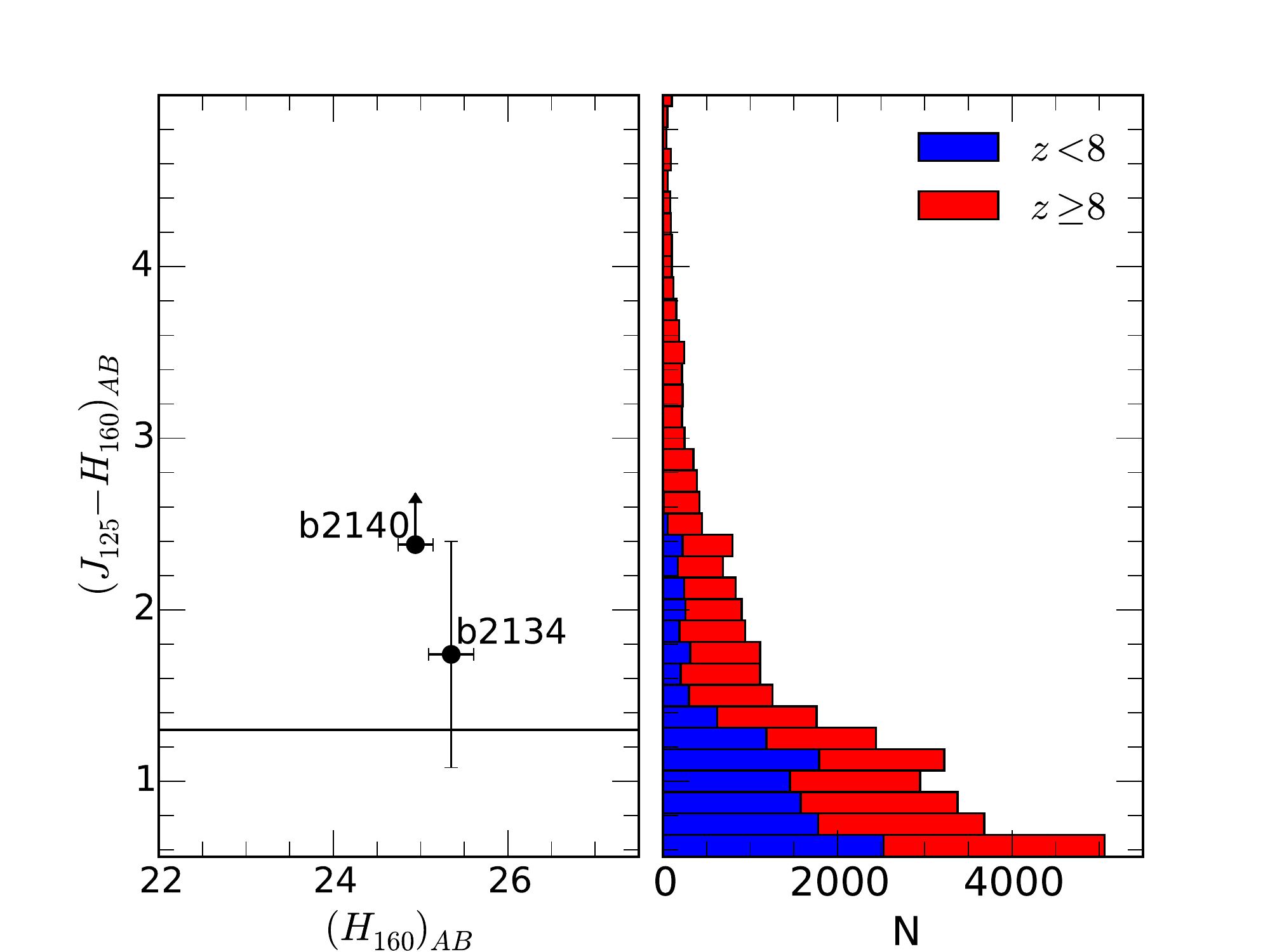}
\caption{Left panel: \jj-\jh\/ dropouts (black filled circles) in the
  \jj-\hh\/ versus \hh-band magnitude plot.  The \jj-\hh\/ color
    is calculated from SExtractor isophotal magnitudes (MAG\_ISO).
  Right panel: distribution of the \jj-\hh\/ color for synthetic
  galaxies at $z>8$ (red) and $z<8$ (blue), showing that the color-cut
  \jj-\hh\/$>1.3$ effectively rejects the large majority of
  contaminants.}
\label{fig:color-mag_z10}
\end{figure}

Our sample of high confidence, high-$z$ candidates consists of five
bright sources detected at $S/N>8$ with inferred redshift peaking at
$z>8$. Two are {\jj-\hh\/ dropouts} ($z\sim 10$) and three are
identified as {\yy-\jh\/ dropouts}, with their most probable redshift
estimated at $z>8.3$. Table \ref{tab:z9-10} contains the photometry
for these sources, Figures \ref{fig:z10} and \ref{fig:z9} show cutout
images centered on each galaxy, the p($z$) distribution, and the best
low and high-$z$ spectral energy distributions (SEDs) fitting the
photometry of the candidates. Finally, the sample reported in this
paper is augmented by four \yy-\jj\/ dropouts, with redshift $z\sim
7.3-8$ (see Table \ref{tab:z8} and Figure \ref{fig:z8}). The IR
color-color selection regions are shown in
Figures \ref{fig:color-color_z8-9} and \ref{fig:color-mag_z10}.

Our alternative search for high-$z$ candidates from the Bayesian
photometric redshifts (Section~\ref{sec:alternative_catalog_z7}) does
not identify additional $z>7$ sources. This provides confidence that
our catalog of bright sources detected at high $S/N$ is not missing
robust candidates, irrespective of the selection technique used.

\subsection{$z\sim 10$ galaxies (\jj-\hh\/  dropouts)}
\label{subsec:candidates_z10}

The selection of the highest redshift galaxies in \BORG\/ relies
primarily on one color (\jj-\hh; see Figure \ref{fig:color-mag_z10}),
associated to non detection in the bluer bands (F350LP and
F105W). F140W is used to verify whether the object is detected in a
second, independent band, and to refine the photometric redshift
estimates. Figure \ref{fig:selection_window_z9-10} shows the expected
redshift distribution of the dropouts for a flat input distribution in
one representative field, obtained through artificial source recovery
simulations (\citealt{oesch2007,oesch2009,oesch2012} and Section
\ref{sec:completeness}). The figure clearly indicates that the color
criteria adopted select sources at $z\gtrsim 9.5$. Our sample consists
of two of them:

{\bf borg\_2134-0708\_774} is a galaxy with magnitude \hh=25.35 and a
very red \jj-\hh\/ color (\jj-\hh$=1.74$). The photometric redshift
probability distribution peaks at $z=10.0$, implying
$M_{AB} \sim -22.2$, albeit there is a broad wing of lower $z$
solutions. The source is clearly resolved and shows extended structure
in both the \jh\/ and \hh-band images. Its intrinsic half-light radius
is $r_e=0\farcs23$, after correction for the broadening introduced by
the Point Spread Function (PSF). Interestingly, the dropout is in
close proximity ($1\farcs46$ center to center) to a foreground galaxy
of magnitude \hh=23.22. The photometric redshift distribution for the
foreground is very broad, but given its compact size, it is likely at
$z\gtrsim 0.5$ and thus can provide at least some gravitational
lensing magnification. We estimated the possible range of
  magnification using the modeling framework developed by
  \citet{baronenugent2015} and \cite{mason2015}. Both methods suggest
  that the magnification is modest ($\mu \lesssim 1.5$) assuming a
  typical mass-to-light ratio, because the foreground galaxy is
  relatively faint. The maximum $\mu\sim 1.5$ is expected if the
  foreground galaxy is at $z\sim 2$, while we predict
  $\mu = 1.2\pm 0.1$ in case of a deflector at $z\sim0.8$, which is the
  redshift at which the lensing optical depth peaks (see
  \citealt{mason2015}).

{\bf borg\_2140+0241\_37} is comparably bright (\hh=24.94) to
borg\_2134-0708\_774, and it is only detected in the two reddest bands
of the survey. Therefore, the photometric redshift has a strong
preference for $z>10$ solutions. Like borg\_2134-0708\_774, this
object is also close in projection ($1\farcs02$ center to center) to a
foreground brighter galaxy (\hh=24.05), also expected to be at
$z\gtrsim 0.5$ because of the compact size, and therefore a potential
lens. The lensing magnification predicted by our modeling is
  $\mu=1.2 \pm 0.1$, essentially identical to that for
  borg\_2134-0708\_774 (see above), because the smaller angular
  separation compensates for the lower luminosity of the lensing
  galaxy.  The dropout galaxy has an extended structure, especially
in the \hh-band image ($r_e=0\farcs37$). Such spatial extension of the
source is larger than that expected for a typical $z\sim10$ candidate
($r_e\lesssim 0\farcs 2$), although still marginally smaller than
confirmed low-$z$ contaminants in CANDELS ($r_e\gtrsim 0\farcs 4$; see
\citealt{holwerda2015}).

\subsection{\yy-\jh\/ dropouts}

The Lyman break selection for \yy-\jh\/ dropouts peaks at $z=8.72$,
with a 95\% confidence region from $z=7.75$ up to $z=9.68$ (see Figure
\ref{fig:selection_window_z9-10}). We identify three candidates with
postage stamps and p($z$) shown in Figure \ref{fig:z9}, all with the
most likely redshift at or below the theoretical median of the
distribution. However, not only is the sample size very small, but a
skewed distribution is expected when the galaxy LF evolves rapidly
over the redshift range covered by the selection \citep{munoz2008}.

{\bf borg\_0116+1425\_630} This object is exceptionally bright for a
$z>8$ candidate (\hh=24.53 mag, corresponding to $M_{AB}\sim -22.8$).
It is detected at high $S/N$ ($S/N_{140}=12.6$, $S/N_{160}=16.1$) and
has a best photometric redshift solution $z=8.4$ and compact size
($r_e=0\farcs 17$ after accounting for the \hh\/ PSF).  This source is
about 0.5 mag brighter than the four $z \ge 7$ candidates in the EGS
field presented by \cite{roberts-borsani2015}, one of which shows an
emission line consistent with Ly$\alpha$ at $z=8.68$
(\citealt{zitrin2015}; and two others have Ly$\alpha$ emission at
$z=7.73$ and $z=7.48$; see
\citealt{oesch2015,roberts-borsani2015}). Our new source thus appears
an ideal candidate for spectroscopic follow-up, with the potential to
elucidate how galaxy formation proceeds for the brightest sources well
into the epoch of reionization. The photometric redshift for the
galaxy shows two peaks (Figure \ref{fig:z9}), but the lower redshift
($z\sim 1.8$) early-type SED is disfavored by the current data and by
the compact size (see \citealt{holwerda2015}). Possibly, part of the
emitted flux of such a bright source could hint at the presence of an
active galactic nucleus \citep{oesch2014}.

{\bf borg\_0956+2848\_85} is the \yy-\jh\/ candidate with the
highest photometric redshift solution ($z=8.7$). Despite being
relatively faint (\hh=26.41) it is confidently detected because of the
long exposure times (e.g., 4400s in \hh). The \yy-\jh\/ drop is also
the most prominent in the sample (\yy-\jh$=2.1$). Finally, its compact
size ($r_e=0\farcs08$ after accounting
for the \hh\/ PSF) strengthens the rejection of the alternative
(already disfavored) photometric redshift solution at $z\sim 1.8$.

{\bf borg\_2229-0945\_548} has a very significant drop in F105W
(\yy-\jh=2.04), which leads to a photo-$z$ distribution sharply peaked
at $z=8.4$. The dropout galaxy, which has \hh= 25.12, is in very close
proximity to a brighter $z\sim 2.1$ passive galaxy (\hh=22.58; center
to center distance equal to 1\farcs48, see bottom panel in Figure
\ref{fig:z9}). Based on modeling of the foreground deflector following
\citet{baronenugent2015} and \cite{mason2015}, we estimate a
gravitational lensing magnification of the dropout flux
$\mu\sim1.3\pm0.3$ or $\mu\sim1.9\pm0.7$, respectively.

\subsection{\yy-\jj\/ dropouts}

While we optimized \BORG\/ for searching objects at $z>8$, the
multi-band nature of the survey allows us to augment the sample of
$z\sim 7-8$ objects as well. We identify candidates in this redshift
range as \yy-\jj\/ dropouts. Four galaxies satisfy the selection
requirements, with a wide range of luminosities ($m_{AB}\sim 25-26.1$
in the \hh-band). As shown in Figure \ref{fig:z8}, the best
photometric redshift solutions lie in the range $7.3 <z<8.0$.  One
caveat to this selection is that, unlike the earlier BoRG[$z$8]
survey, which relied on the medium-band filter F098M for {\it Y}-band
imaging, the use of F105W implies a partial overlap with F125W,
resulting in a potentially higher contaminant fraction for \yy-\jj\/
LBGs \citep{stanway2008}. However, the analysis of the size
distribution of the sample appears reassuring because all four objects
have $r_e\leq 0\farcs 25$, and only one has $r_e>0\farcs2$ after
accounting for the PSF shape. Thus we have confidence that even in
absence of longer wavelength observations and despite the partial
overlap of F105W with F125W, the purity of the sample that we
constructed is high (see Section \ref{sec:contamination}).

\section{Number density and luminosity function of bright LBGs at $z\gtrsim8$}
\label{sec:LF}

One of the key science drivers of \BORG\/ is to characterize the
number density of bright ($L>L_*$) galaxies at $z>8$. In the initial
25\% of the survey, which is presented in this paper, we identified
bright, but rare candidates (five in total at $z\sim 8.5$ and
$z\sim10$; see Section \ref{sec:candidates}). These detections
translate into preliminary limits of the galaxy number density and UV
LF after quantifying the completeness of our search and estimating of
the contamination rate.

\subsection{Completeness and selection functions}\label{sec:completeness}
\begin{figure*}[]
\center
\includegraphics[scale=0.8]{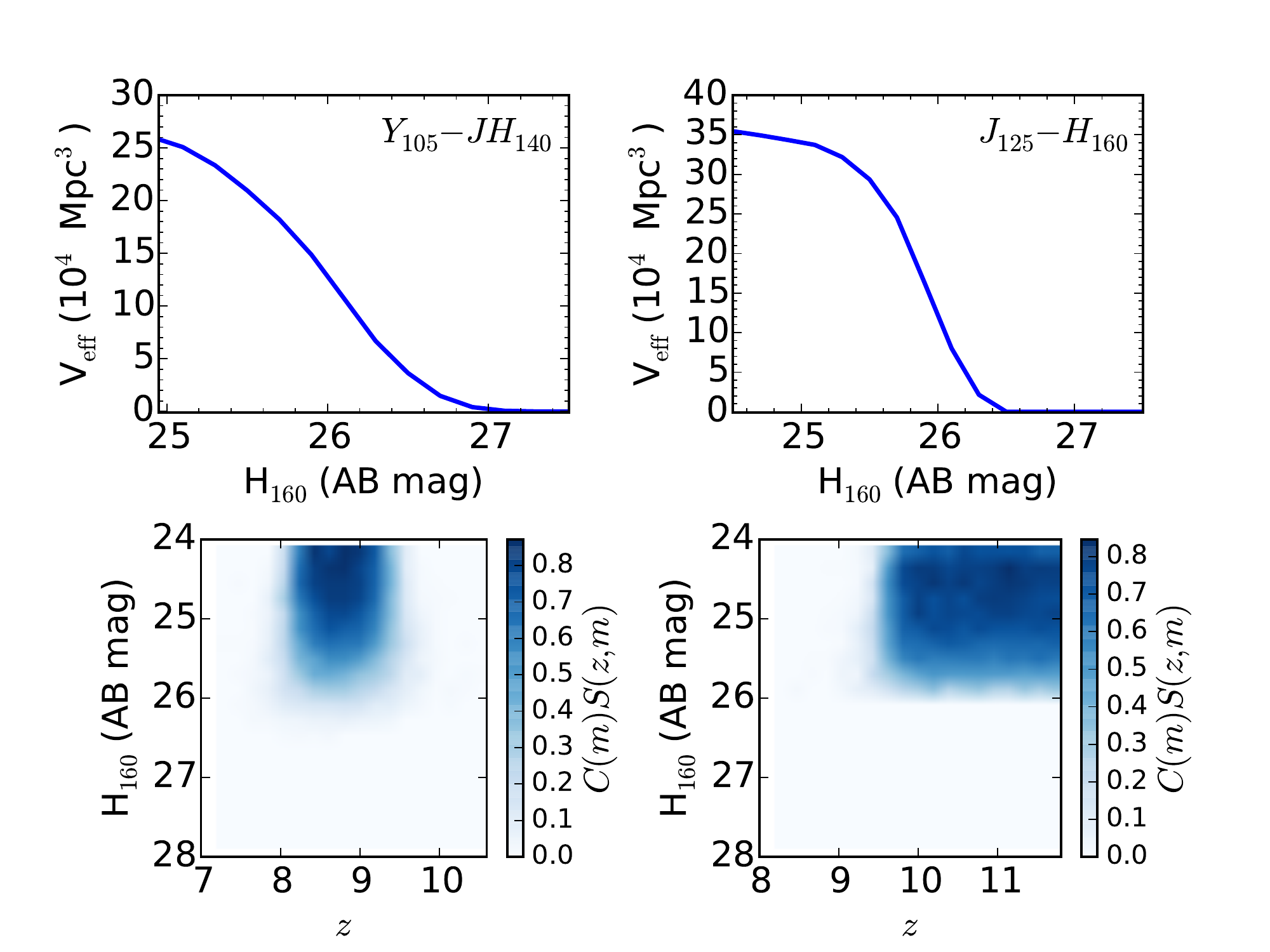}
\caption{Top panels: Effective comoving volume $V_{\rm eff}$ in Mpc$^3$ as
  a function of the \hh-band magnitude for our selection of LBGs as
  \yy-\jh\/ (left panel) and \jj-\hh\/ dropouts (right panel). Bottom
  panels: Plot of the selection function $S(z,m)$ for the \yy-\jh\/
  (left panel) and \jj-\hh\/ dropout (right panel) samples obtained
  for the field borg\_0116+1425 representative of a typical \BORG\/
  pointing.  $S(z,m)$ was derived from simulations, recovering
  artificial sources inserted in the images
  \citep{oesch2007,oesch2009,oesch2012}.}
\label{fig:V_eff}
\end{figure*}

Following the prescription of \cite{oesch2007,oesch2009,oesch2012}, we
ran simulations of artificial source recovery to derive the
completeness function, $C(m)$, and magnitude-dependent redshift
selection function, $S(z, m)$, at $z\sim9$ and $z\sim10$ for each
\BORG\/ field. A detailed discussion of the simulations is presented
by \citet{oesch2012} (see also \citealt{bradley2012,schmidt2014} for
previous applications of the method to the BoRG[$z$8] survey). To
summarize the method, artificial sources with a range of input
magnitudes, sizes, spectral energy distributions, and redshifts are
added to the science images. Then dropout catalogs are constructed
following the steps described in Sections \ref{sec:catalogs} and
\ref{sec:selection}. From these catalogs we construct the completeness
$C(m)$ and source selection $S(z, m)$ functions. The procedure is
carried out for each individual field (see Figure \ref{fig:V_eff} for
borg\_0116+1425 field, representative of all those in our
dataset). The sum of all completeness weighted selection functions,
integrated over redshift, then determines the effective comoving
volume probed as a function of source brightness:
$$V_{\rm eff}(m)=\int_0^{\infty} S(z,m)C(m)\frac{dV}{dz}dz$$ 
$V_{\rm eff}(m)$ takes into account all aspects of the selection of
high-$z$ sources, including (1) loss of volume due to foreground
sources and/or areas affected by persistence; (2) decreased detection
efficiency as the survey magnitude limit is approached; (3) effects of
photometric scatter which can move artificial sources at high-$z$
outside the parameter space for selection. $V_{\rm eff}(m)$ is shown
in Figure \ref{fig:V_eff}, from which it is immediately clear that the
recovery efficiency of \BORG\/ drops around $m_{AB}=26.2$ in the
\hh-band for sources detected at $S/N\geq 8$. At bright magnitudes,
the effective volume of the search for the $z\sim 9$ and $z\sim 10$
samples is respectively $(2.5-3.5)\times 10^5$ Mpc$^3$ comoving. 

\subsection{Contamination}\label{sec:contamination}

Catalogs of high-$z$ candidates selected from broadband imaging are
potentially affected by contamination of low-$z$ interlopers with
similar colors, which may be Galactic stars, low/intermediate redshift
galaxies, or spurious sources \citep{bouwens2015_10000gal}. 

In general stellar contamination is not a significant source of
concern for resolved sources detected at $S/N\gtrsim10$, since in that
case the {\tt SExtractor} CLASS\_STAR parameter can reliably
discriminate between stars and galaxies
\citep{bouwens2015_10000gal}. All sources in our  main samples
have CLASS\_STAR$<0.5$  (see Table \ref{tab:z9-10}), which is
comfortably different from CLASS\_STAR$>0.9$ typically measured for
point sources. In addition, stars with red IR colors have a surface
density that remains relatively flat in the magnitude range
$21 \lesssim m_{AB} \lesssim 26$ \citep{holwerda2014}.

The color-color criteria we adopted for the selection of \yy-\jh\/ and
\jj-\hh\/ dropouts exclude the contamination from dwarf stars based on
the colors of these sources.  On the other hand, the sample of
\yy-\jj\/ dropouts is potentially contaminated by cold red stars.
From archival near-IR {\it HST} datasets
\citep{ryan2011,holwerda2014}, we estimate that $n\sim 2$ T-dwarf
stars may enter our selection box within the current survey area. From
the values of CLASS\_STAR and $r_e$ for \yy-\jj\/ dropouts we conclude
that only borg\_0835+0310\_145 is compact enough to be a
potential stellar interloper of the $z\sim 7.5$ dropout sample.

We do not expect that contamination by spurious sources is a concern
because of our requirement of $S/N>8$ in the detection images combined
with stringent color cuts. \citet{schmidt2014} analyzed BoRG[$z$8]
data to characterize the noise distribution, and found that on
average, one spurious source for each WFC3 pointing is detected at
$S/N>8$ because of hot pixels or detector persistence.  However, the
colors of these spurious sources are not as red as $z>8$ dropouts
because hot pixels or persistence affect all IR bands. In fact,
\citet{schmidt2014} found that no spurious source was identified as a
dropout within the $350$ arcmin$^2$ of the full BoRG[z8]
dataset. Applying the conclusions of that study to the \BORG\/ data
analyzed here, we expect $n\lesssim 0.3$ spurious sources in our
samples. Note also that since images in all filters are acquired for
each pointing in a single visit (lasting less than 8 hours),
contamination by transients such as supernova events is negligible
since only a $z>8$ event would have the right colors to enter into our
selection. 

Finally, we discuss the most significant source of contamination:
foreground galaxies with colors similar to $z>8$ sources. Two main
classes of objects may enter our LBG selection: intermediate-age
galaxies at $z\sim 1-3$ with a prominent Balmer break, and strong line
emitters, which have IR broadband flux dominated by nebular lines
(e.g. H$\alpha$, \oo2, and \o3; \citealt{atek2011,pacifici2015}) and
an undetected faint continuum flux at optical wavelengths
\citep{bouwens2015_10000gal}. Spectroscopic follow-up of $z\sim 8$
dropouts in BoRG[$z$8] has found no evidence of strong emitters after
targeting 15 sources \citep{treu2012,treu2013,baronenugent2015}
selected from $\sim 350$ arcmin$^2$. Even more stringently,
\cite{bouwens2015_10000gal} estimate that the extreme line emitters
capable of contaminating a dropout sample at $z\sim 8$ have a density
of $\sim 10^{-3}$ arcmin$^{-2}$. We thus estimate that extreme line
emission is not a significant concern, although one caveat is that the
only $z\gtrsim 10$ candidate identified in UDF field may be an \o3\/
emitter at $z\sim2.2$ (see \citealt{brammer2013}). The primary source
of contaminants in our sample are thus expected to be intermediate
redshift galaxies with a strong Balmer break. Estimating the precise
contamination fraction from these sources is very challenging, but
 we expect it to be in the range of $\sim 30$\% based on the
  analysis of spectral templates. This figure is comparable to
  previous estimates of the BoRG[$z$8] sample purity at the $20-30\%$
  level \citep{trenti2011,bradley2012,schmidt2014}. This rate is
  larger than the typical contamination rate $\lesssim 10\%$ found in
  Lyman-break samples from legacy fields such as HUDF and CANDELS,
  that have extensive multi-observatory coverage
  \citep{bouwens2015_10000gal}.  In the absence of observations with
Spitzer/IRAC which would help to discriminate between intermediate-$z$
ellipticals and $z>8$ starbursts, it is possible to use size as a
proxy for the \hh-[4.5] color \citep{holwerda2015}, and reassuringly
our dropouts sources are generally too compact to be
contaminants.  One caveat is that the size proxy is unable to
  discriminate against contamination from sources with unusual colors
  at $z\sim 6-7$, since they would be similarly compact. Of course,
  the availability of observations over a wider range of wavelengths
  would help to identify the presence, if any, of such population in
  our samples.

The presence of a non-negligible, but not overwhelmingly large
contamination fraction can be inferred from the analysis of the
redshift probability distribution for the candidates in our sample,
 as well. For sources at $z\sim 10$, we measure that the
probability of being at $z<8$ is $p=39$\%, while for $z\sim 9$
dropouts the photo-$z$ estimates are more peaked at high redshift, and
there is only an average probability $p=11$\% for the candidates to be
at $z<7$. These estimates are in agreement with the study by
\cite{pirzkal2013}, where a $\sim21$\% contamination fraction is
derived for typical samples of galaxies at $8<z<12$ identified from
{\it HST} imaging.

Combining all the different approaches to the contamination issue, we
assume 30\% as  baseline estimate of the contamination rate, 
  which is close to the weighted average from the photo-$z$
  [$(0.39\times 3 + 0.11\times 2)/5 \sim 0.28$]. Thus, we would expect
that one to two of the five sources reported in this paper may be
low-redshift interlopers.

\subsection{Determination of the luminosity function}

\begin{figure}[!t]
\center
\includegraphics[scale=0.45]{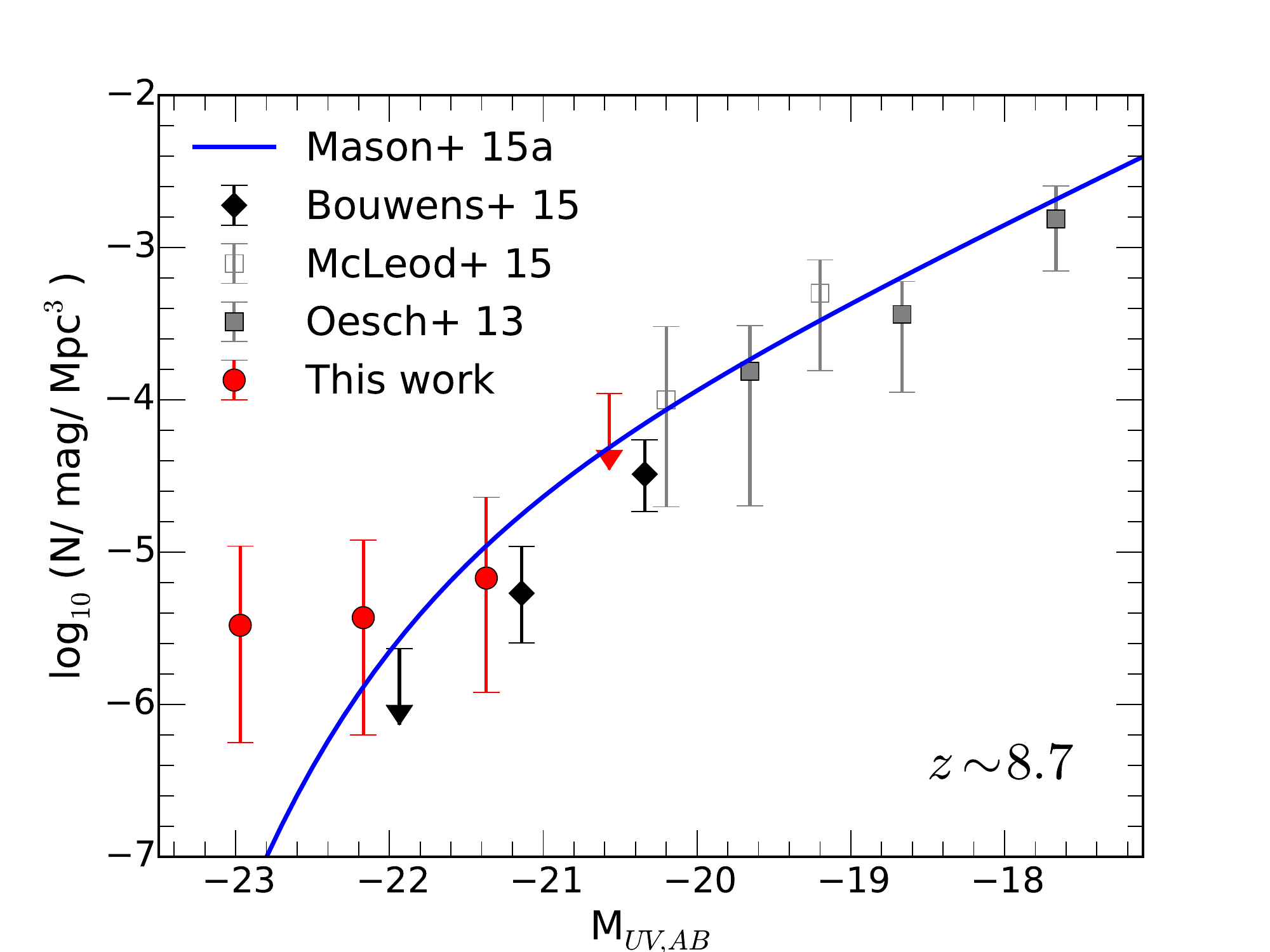}
\includegraphics[scale=0.45]{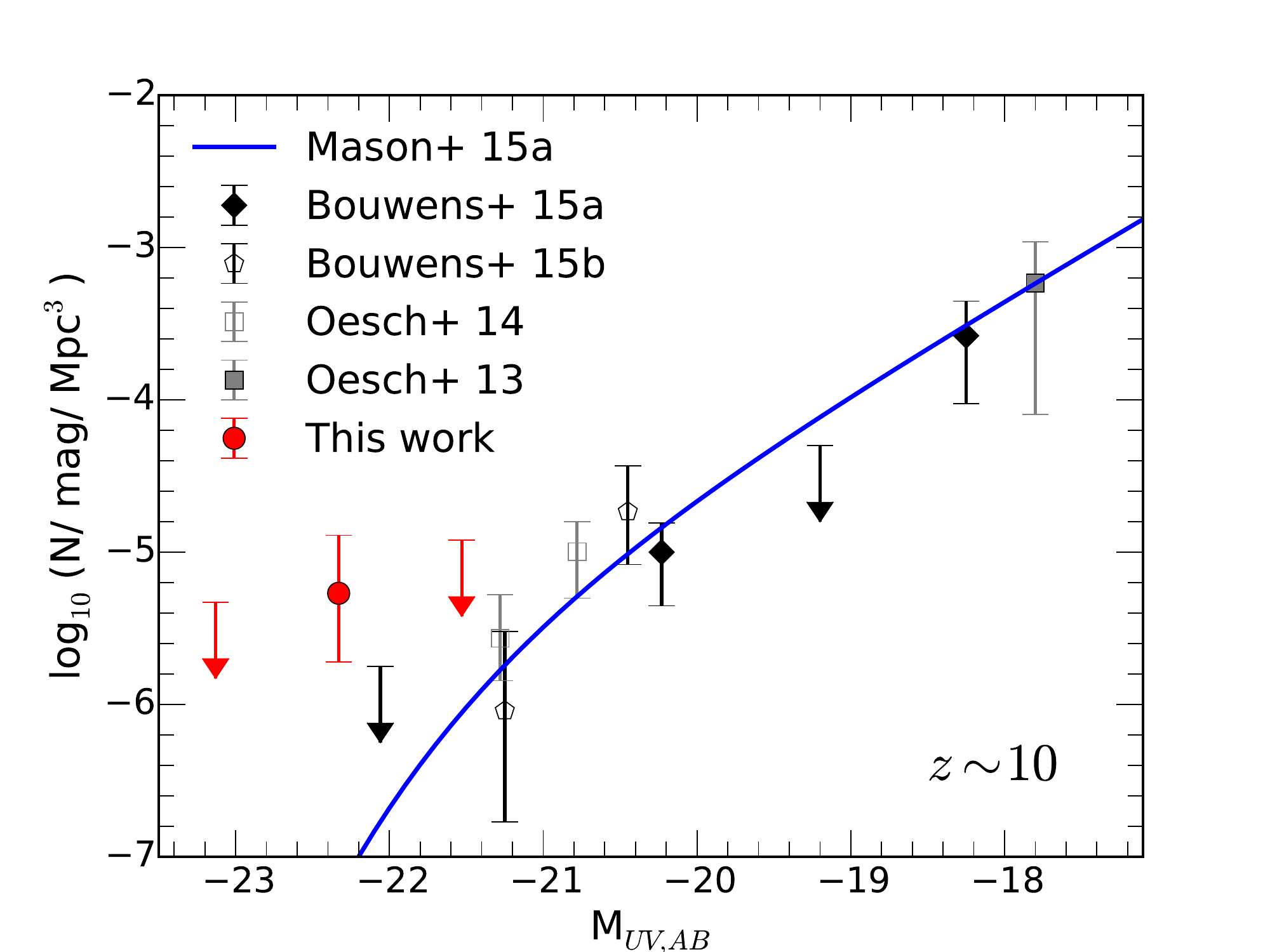}
\caption{Step-wise determinations of the UV LF at $z \sim 8.7$ (top
  panel) and $z\sim 10$ (bottom panel),  without magnification
    lensing corrections. The red filled circles and upper limits
  refer to this paper, other symbols to
  \cite{oesch2013,oesch2014,bouwens2015,bouwens2015_10000gal,mcleod2015}
  as labeled.  The over-plotted blue line indicates the galaxy UV
  luminosity function as from \cite{mason2015_LF}.}
\label{fig:LF}
\end{figure}

Combining the effective volume and contamination estimates, we
derive a step-wise LF for the $z\sim 9$ and $z\sim10$ samples, which
we report in Table~\ref{tab:stepwiseLF} and plot in
Figure \ref{fig:LF}. The determination is severely limited by the
large Poisson uncertainty, but the comparison with existing
constraints on the LF, shown as gray points in the Figure, is
informative. Our determination of the bright end of the $z\sim 9$
luminosity function is consistent with the latest measurement by
\citet{bouwens2015} at $M_{AB}>-21.5$, and at $M_{AB}=-22.2$ the
measured number density of $3.7^{+8.3}_{-3.1}\times 10^{-6}
\mathrm{Mpc^{-3} mag^{-1}}$ is within the predictions of the LF model
of \citet{mason2015_LF}, which is  successful in describing the LF
evolution with redshift. The striking difference with previous
searches and with theoretical predictions is the detection of the
exceptionally bright candidate borg\_0116+1425\_630, which if
confirmed, would argue against an exponential decline at the
bright end and point instead of a power-law LF. One possibility is, of
course, that such object is a contaminant, but intriguingly the recent
spectroscopic confirmation at $z\sim8.7$ of a $m_{AB}\sim 25$ source
\citep{zitrin2015} may suggest that intense starbursts of the order
of $50~\mathrm{M_{\odot} yr^{-1}}$ could become relatively more common
as the redshift increases, and merger driven activity increases over
smooth gas accretion.

At $z\sim 10$, the situation is similar. At face value, our LF
determination appears too high compared to expectations for objects
with $m_{AB}<25.5$. One possibility to explain our results would be
significant lensing magnification, because both candidates are close
in proximity to brighter foreground sources. At the current stage and
without follow-up studies of the two candidates to increase the
confidence on their $z\sim 10$ nature, it is difficult to draw firmer
conclusions. Of course, the full \BORG\/ dataset will allow us to
investigate whether this initial overabundance of very bright
candidates is confirmed or not and to systematically account for the
effect of lensing magnification on the LF via Bayesian methods
\citep{schmidt2014,mason2015}.

As an additional consistency check for the number of detections
reported here compared to theoretical expectations, we estimated the
number density by integrating the \citet{mason2015} LF model (see also
\citealt{trenti2010,tacchella2013}) over the effective volume of the
survey. For $z\sim9$, we estimate a total of $\langle n
\rangle=1.1^{+1.5}_{-0.7}$ detections, consistent with the $n=2.1$
observed after accounting for $30\%$ contamination. For $z\sim 10$,
the expectation is $\langle n \rangle=0.1^{+0.25}_{-0.08}$ detections,
so this is in mild ($\sim 2\sigma$) tension with the observed number
after accounting again for $30\%$ contamination. If compared to
expectations from the $z\sim 8$ LF, which would predict $\langle n
\rangle \sim 11$ detections in the survey, our result of an
contamination-corrected sample size of $n \sim 3.5$ sources indicates
a decline in bright galaxies with increasing redshift, confirming the
clear trend previously established observationally and by theoretical
modeling (e.g.,
\citealt{oesch2012,bouwens2015_10000gal,mason2015_LF}).

Finally, regarding the sample of four objects at $z\sim7.3-8$, we
defer the study of the LF until the full dataset has been acquired,
since the new area ($\sim 130$ arcmin$^2$) is only a modest
improvement over the existing BoRG[$z$8] data ($\sim 350$ arcmin$^2$)
and the determination from the combination of all {\it HST} archival
data ($\sim 1000$ arcmin$^2$; \citealt{bouwens2015}).


\section{Conclusion}
\label{conclusion}

In this paper we presented the design and initial results of the
\BORG\/ survey, a large (480 orbits) pure-parallel imaging program
with {\it HST}/WFC3 which is acquiring medium depth ($m_{AB}\sim 27$,
$5\sigma$ point source), random pointing imaging at optical and
infrared wavelengths over a total of $\sim 550$ arcmin$^2$, divided
among more than 100 independent lines of sight (GO13767, PI Trenti). 
The primary goal of \BORG\/ is the detection of $L>L_*$ galaxy
candidates at $z>8$ 
taking advantage of the large number of independent lines of sight to
minimize the impact of cosmic variance. 

The key results are the following:
\begin{itemize}

\item Through an optimized design of the pure-parallel opportunities
  (Section \ref{sec:survey}), we have been able to achieve an image
  quality nearly equivalent to that of primary dithered imaging
  (Section \ref{sec:quality} and Figure \ref{fig:pp_vs_dith}).

\item From the initial $\sim 25\%$ of the survey, we identified five
  very bright galaxy candidates at $z>8.3$ with $m_{AB}\sim 24.5-26.5$
  detected at $S/N>8$, contributing significantly to increase the
  small sample sizes of intrinsically bright objects identified at
  $z>8$ from legacy fields \citep{bouwens2015,bouwens2015_10000gal}. To select
  the objects we resorted to a combination of color selection (LBG
  technique; see \citealt{steidel1996}) and Bayesian photometric
  redshift estimates, with the candidate properties presented in
  Section \ref{sec:candidates} and in
  Figures \ref{fig:z10} and \ref{fig:z9}.

\item One source, borg\_0116+1425\_630, is exceptionally bright, with
  \hh=24.5 and best photometric redshift $z=8.4$, implying
  $M_{AB}=-22.75$ and a star formation rate (SFR) of $\sim 66\,
  \rm{M_{\odot} yr^{-1}}$.  The existence of such a bright source at
  $z>8$, if confirmed, would be a major indication that galaxy
  formation at early times can undergo phases of rapid bursts with
  sustained SFR, possibly because of frequent major mergers.

\item After accounting for completeness corrections and for a modest
  contamination rate (Section \ref{sec:contamination}), the inferred
  number density is consistent with previous observations and
  theoretical modeling (e.g.,
  \citealt{bouwens2015_10000gal,mason2015}), although our brightest
  bins at $z\sim 9$ and $z\sim 10$ have an excess compared to
  expectations (Figure \ref{fig:LF}). However, the excess might be
  explained with small number fluctuations, since the number counts
  predictions are not significantly different. For example, we observe
  three sources at $z\sim 9$ versus the expectation of $\langle n
  \rangle=1.1$. One additional contributing factor could be
  line-of-sight gravitational lensing since some of our candidates are
  in close proximity with brighter and more massive foreground sources
  (see \citealt{mason2015,baronenugent2015}).

\end{itemize}

Overall, our new candidates, and in particular borg\_0116+1425\_630,
are ideal for follow-up observations either with Spitzer/IRAC to
measure the rest-frame optical light  and improve sample
  purity\footnote{Our team has been awarded time for follow-up of the
    $z\sim 9-10$ candidates presented in this paper after submission
    of the manuscript. These Spitzer observations (Program 12058, PI.
    R. J. Bouwens) will be carried out by the end of 2016.}, or with
near-IR spectroscopy. The prospects for the latter to succeed are
favorable based on the recent report of detection of Ly$\alpha$ in a
comparably bright source by \citealt{zitrin2015}.

The evolution of the bright end of the galaxy LF into the epoch of
reionization is an active topic of research, with potential to
elucidate when and how the feedback processes that affect the
brightest galaxies at lower redshift come in place. The current status
of the field is somewhat unclear, with some tension between
ground-based determinations at $z\sim 7$, which suggest a power-law
fall-off rather than a Schechter function at the bright end
(\citealt{bowler2014}), while $z\sim 8$ data from {\it HST} prefer an
exponential cut-off, including the large area determination from our
previous BoRG[$z$8] survey (\citealt{schmidt2014}; but see
\citealt{finkelstein2015} for support to a comparably good fit with a
single power law). Our preliminary results reported here at $z>8$ are
not conclusive, but the full dataset which is being acquired will
improve the constraints with an increase of a factor four in
area. Finally, when combined with other ongoing programs that will
find bright (lensed) high-$z$ candidates, such as GLASS (GO13459,
PI. T. Treu, \citealt{schmidt2014GLASS,treu2015}) and RELICS (GO14096,
PI. D.  Coe), \BORG\/ will contribute to create a legacy set of
near-IR observations from which the most promising spectroscopic
targets for early {\it JWST}/NIRSPEC observations can be selected.

\acknowledgments{We thank Camilla Pacifici and Robert Barone-Nugent
  for useful discussions and the anonymous referee for valuable
  comments.  This work was partially supported by grants {\it HST}/GO
  13767, 12905, and 12572.}

Facilities: \facility{{\it HST}(WFC3)}.

\bibliographystyle{apj}

\bibliography{valebib} 
\appendix 
\section{Appendix A- Other possible high-$z$ candidates}

In this Appendix we briefly investigate and present results on the
search for high-$z$ sources in the dataset based on alternative
criteria. In addition, we report the candidates that passed the Lyman
break color selection for $z>8$ sources, but were removed from the
main sample because their Bayesian photometric redshift indicated a
best solution at $z<7$. Overall, we discuss six additional high-$z$
candidates, with four of them being possible/likely contaminants
because of a preferred low-$z$ solution in the photometric redshift
distribution. Four of the six sources arise from the use of F160W as
sole detection image (rather than the combination of F140W and
F160W). One candidate is excluded from the main sample because of
the low-$z$ peak in $p(z)$. Finally, one additional candidate is
identified by adopting the \citet{bouwens2015_10000gal} selection for
$z\sim 8$ sources (but would have been excluded from the main sample
in any case because of a preferred low-$z$ photometric redshift
solution).

\subsection{Additional candidates from F160W-only catalogs}

Catalogs constructed using F160W-only as detection image have been
processed to search for candidates at $z>7$ following the main text
analysis based on F140W+F160W selection. We identified three new
sources at $z>7$, that were not included in the main samples because
of slightly different colors or because a hot pixel was present within
their segmentation maps. Interestingly, one of these new candidates,
borg\_1209+4543\_1696, is identified as a \yy-\jh\/ dropout, with a
photometric redshift distribution strongly peaked at $z=8.56$ and
\hh$\sim 25.0$.  The reason why the source is not in the main catalog
is because of the presence of an hot pixel, correctly identified by
the reduction pipeline, in the outskirts of the dropout image. The
pixel is included in the segmentation map of the F160W+F140W detection
image, thereby excluding the galaxy from further analysis owing to our
choice to discard any source with one or more pixels having zero
weight (infinite rms). With the selection in F160W only, the galaxy
segmentation map avoids inclusion of the hot pixel, and {\tt
  SExtractor} photometry is performed without incurring in unbound
errors. Since the defect is located in the outer parts of the image of
the galaxy, we consider this source a credible high-$z$ candidate,
albeit its relatively large size would indicate the possibility that
the low-$z$ solution is preferred. The other two sources are less
interesting and have a high probability of being contaminants, with
colors near/at the boundary of the selection region.

\subsection{Search for sources with Bayesian photometric redshift at
  $z>7$}

We analyzed the photometric redshift probability distribution for each
galaxy in our catalogs, constructed following the procedure introduced
in Section \ref{sec:photo-z}. No new candidates at $z>7$ have been
identified this way.

\subsection{Candidates with degenerate photo-$z$ solutions}

Three sources listed in Table \ref{tab:additional_sources} and shown
in Figure \ref{fig:additional_candidates} have a p($z$) distribution
peaked at $z\sim 1-2$ and magnitudes \hh$\sim 25-26.5$. The existing
data make it difficult to evaluate whether this is an effect of
photometric scatter versus the presence of a population of interlopers
with similar near-IR colors, especially because the compact effective
radii of these sources would corroborate the high-$z$ photometric
solution.
 
\subsection{ \yy-\jj\/ selection following \cite{bouwens2015_10000gal}} 

The selection of $z\sim 7.5-8$ sources in Section~\ref{sec:selection}
has been optimized for the \BORG\/ filter set. Here, we select sources
following the criteria adopted for legacy fields data by
\cite{bouwens2015_10000gal}, with the main difference being the availability of
a larger number of blue (non-detection) bands from surveys such as
GOODS/CANDELS \citep{grogin2011}. The selection criteria are: 

$$S/N_{350}<1.5$$ 
$$S/N_{125}\ge6$$ 
$$S/N_{140}\ge6$$
$$S/N_{160}\ge4$$
$$m_{105}-m_{125}>0.45$$
$$m_{105}-m_{125}>0.75\cdot(m_{125}-m_{160})+0.525$$
$$m_{125}-m_{160}<0.5$$

\begin{figure*}[!h]
\center
\includegraphics[scale=0.48]{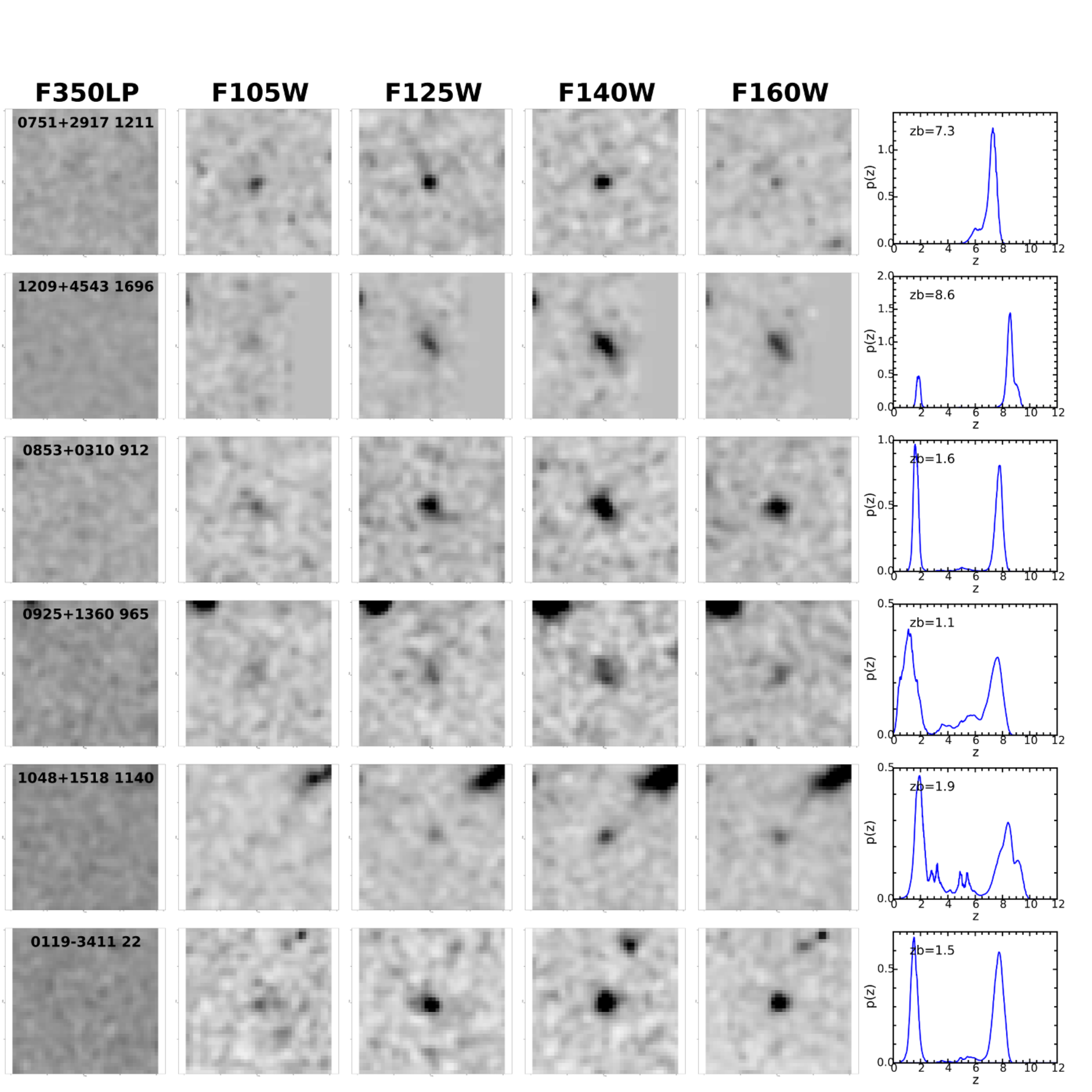}
\caption{Postage-stamps cutouts and photometric redshift probability
  distribution p($z$) for the additional dropout candidates discussed in the
  Appendix. }
\label{fig:additional_candidates}
\end{figure*}

With this selection, we identified one additional \yy-\jj\/ dropout
borg\_0119-3411\_22, which is quite bright (\hh=25.06). In the
near-IR color-color space, the source lies close to the boundaries of
the selection box for high-$z$ objects. Therefore, it is not
surprising that the photometric redshift distribution shows both a low
and an high redshift peak (Figure \ref{fig:additional_candidates}), with the
low-$z$ solution marginally preferred. The uncertainty in the nature
of this source is compounded by the relatively short exposure time in
F350LP which was constrained by readout conflicts with the primary
observations associated to this opportunity. 

The conclusion from these additional searches is that the failure to
identify credible additional candidates both with an alternate color
selection, and with a photo-$z$ analysis (Section \ref{sec:photo-z}),
confirms that the selection criteria adopted in the main text are
robust.

\clearpage

\begin{turnpage}
\begin{deluxetable*}{lccccccccccccccc}
\tabletypesize{\footnotesize}
\tablecolumns{14}
\tablewidth{0pt} 
\tablecaption{\BORG\/ fields coordinates, exposure times, and limiting magnitudes}
\tablehead{\colhead{Field ID} & \colhead{$\alpha$(J2000)} & \colhead{$\delta$(J2000)} & \colhead{\# of}  & \colhead{E(B-V)}  & \multicolumn{2}{c}{F350LP} & \multicolumn{2}{c}{F105W} &\multicolumn{2}{c}{F125W} & \multicolumn{2}{c}{F140W} & \multicolumn{2}{c}{F160} & \colhead{Effective}\\
        \colhead{}& \colhead{(deg)} & \colhead{(deg)}& \colhead{orbits} && \colhead{exp time}  & \colhead{m$_{\rm lim}$} & \colhead{exp time} & \colhead{m$_{\rm lim}$} & \colhead{exp time} & \colhead{m$_{\rm lim}$} & \colhead{exp time} & \colhead{m$_{\rm lim}$} & \colhead{exp time} & \colhead{m$_{\rm lim}$} & \colhead{area}}
\startdata
borg\_0132+3035$^\ast$  &   23.11 & +30.59 & 4  & 0.042 & 1880 &       & 2109 &       & 2109 &       & 1809 &       & 2409 &       &     \\
borg\_0133+3043$^\ast$  &   23.37 & +30.72 & 4  & 0.035 & 1830 &       & 2159 &       & 2109 &       & 1759 &       & 2409 &       &     \\
borg\_0134+3034$^\ast$  &   23.49 & +30.58 & 4  & 0.037 & 1860 &       & 2159 &       & 2109 &       & 1759 &       & 2409 &       &     \\
borg\_0134+3041$^\ast$  &   23.43 & +30.68 & 4  & 0.035 & 1830 &       & 2159 &       & 2109 &       & 1156 &       & 1606 &       &     \\
borg\_0116+1425        &   19.06 & +14.41 & 4  & 0.035 & 2095 & 26.99 & 2209 & 26.44 & 2059 & 26.57 & 1759 & 26.45 & 2409 & 26.45 & 4.40\\
borg\_0119-3411        &   19.68 & -34.18 & 3  & 0.023 & 1306 & 26.70 & 1606 & 26.23 & 1506 & 26.17 & 1306 & 26.55 & 1759 & 26.34 & 4.27\\
borg\_0337-0507        &   54.38 &  -5.12 & 4  & 0.038 & 1967 & 26.95 & 2109 & 26.41 & 2059 & 26.49 & 1709 & 26.60 & 2409 & 26.46 & 4.42\\
borg\_0554-6005        &   88.39 & -60.09 & 5  & 0.049 & 2252 & 27.05 & 2512 & 26.42 & 2412 & 26.81 & 2059 & 26.90 & 2812 & 26.70 & 4.12\\
borg\_0751+2917        &  117.71 & +29.28 & 5  & 0.037 & 2210 & 26.95 & 2462 & 26.59 & 2412 & 26.60 & 2009 & 26.72 & 2812 & 26.55 & 4.43\\
borg\_0853+0310        &  133.18 &  +3.16 & 3  & 0.043 & 1392 & 26.93 & 1556 & 26.60 & 1506 & 26.50 & 1256 & 26.59 & 1709 & 26.38 & 4.49\\
borg\_0925+1360        &  141.31 & +14.00 & 3  & 0.027 & 1510 & 26.88 & 1706 & 26.55 & 1506 & 26.48 & 1306 & 26.43 & 1859 & 26.33 & 4.52\\
borg\_0925+3439        &  141.33 & +34.65 & 4  & 0.017 & 2039 & 27.03 & 2159 & 26.58 & 2059 & 26.57 & 1759 & 26.61 & 2459 & 26.51 & 4.47\\ 
borg\_0953+5157        &  148.26 & +51.95 & 4  & 0.008 & 1809 & 27.24 & 2359 & 26.96 & 2309 & 26.83 & 1959 & 26.91 & 2662 & 26.67 & 4.44\\
borg\_0956+2848        &  149.10 & +28.80 & 7  & 0.016 & 2940 & 27.15 & 3865 & 26.77 & 3768 & 26.78 & 3215 & 26.88 & 4418 & 26.76 & 4.43\\ 
borg\_1015+5945$^\dagger$&  153.74 & +59.75 & 3+4& 0.009 & 3084 &       & 4215 &       & 4018 &       & 3468 &       & 4718 &      &      \\
borg\_1018+0544        &  154.47 &  +5.74 & 4  & 0.017 & 2000 & 27.07 & 2109 & 26.61 & 2009 & 26.59 & 1759 & 26.66 & 2409 & 26.52 & 4.44\\
borg\_1048+1518        &  161.97 & +15.30 & 3+4& 0.024 & 2478 & 27.15 & 3112 & 26.89 & 2912 & 26.80 & 2512 & 26.86 & 3518 & 26.68 & 4.43\\ 
borg\_1048+1518$^\dagger$&  161.97 & +15.30 & 4  & 0.024 & 1980 &       & 2059 &       & 1959 &       & 1659 &       & 2309 &       &     \\
borg\_1103+2913        &  165.68 & +29.22 & 3+3& 0.025 & 2575 & 27.21 & 2912 & 26.83 & 2812 & 26.80 & 2312 & 26.87 & 3212 & 26.73 & 4.46\\
borg\_1106+3508        &  166.53 & +35.14 & 5  & 0.016 & 2480 & 27.15 & 2762 & 26.63 & 2662 & 26.76 & 2209 & 26.80 & 3112 & 26.65 & 4.44\\
borg\_1115+2548        &  168.66 & +25.80 & 4  & 0.015 & 2151 & 27.10 & 2462 & 26.80 & 2412 & 26.76 & 2009 & 26.82 & 2762 & 26.63 & 4.50\\
borg\_1127+2653$^\star$ &  171.81 & +26.88 & 3  & 0.015 &      &&&&&&&&&&\\
borg\_1142+3020        &  175.62 & +30.34 & 4  & 0.018 & 2130 & 27.19 & 2159 & 26.83 & 2109 & 26.77 & 1759 & 26.79 & 2409 & 26.61 & 4.51\\
borg\_1152+3402        &  177.91 & +34.03 & 3  & 0.017 & 1154 & 26.88 & 1456 & 26.56 & 1406 & 26.45 & 1156 & 26.57 & 1606 & 26.40 & 4.40\\
borg\_1154+4639        &  178.44 & +46.45 & 6  & 0.028 & 2583 & 27.38 & 3412 & 27.04 & 3212 & 26.90 & 2712 & 27.07 & 3718 & 26.83 & 4.29\\
borg\_1160+0015        &  179.98 & + 0.25 & 3  & 0.028 & 1473 & 26.93 & 1606 & 26.55 & 1506 & 26.48 & 1256 & 26.51 & 1806 & 26.42 & 4.48\\
borg\_1209+4543        &  182.36 & +45.72 & 3+5& 0.012 & 3500 & 27.46 & 3918 & 26.84 & 3718 & 27.10 & 3165 & 27.15 & 4421 & 26.93 & 4.44\\
borg\_1410+2623        &  212.41 & +26.38 & 4  & 0.014 & 2210 & 27.22 & 2462 & 26.69 & 2412 & 26.81 & 2009 & 26.90 & 2812 & 26.66 & 4.44\\
borg\_1438-0142        &  219.45 &  -1.70 & 3+5& 0.037 & 3393 & 27.28 & 3918 & 27.01 & 3768 & 26.96 & 3165 & 27.02 & 4421 & 26.84 & 4.37\\
borg\_1520-2501$^\diamond$& 230.08 & -25.02 & 3  & 0.142 & 1872 & 26.58 & 1356 & 26.31 & 1256 & 26.27 & 1006 & 26.26 & 1506 & 26.18 & 4.28\\
borg\_1525+0955        &  231.17 &  +9.92 & 3  & 0.034 & 1230 & 26.88 & 1456 & 26.57 & 1356 & 26.48 & 1156 & 26.51 & 1609 & 26.29 & 4.31\\
borg\_1525+0960        &  231.19 & +10.00 & 5  & 0.033 & 2154 & 27.15 & 2462 & 26.84 & 2362 & 26.77 & 2009 & 26.83 & 2862 & 26.64 & 4.23\\
borg\_2134-0708        &  323.54 &  -7.13 & 3+4& 0.028 & 3605 & 26.90 & 3715 & 26.27 & 3515 & 26.45 & 2965 & 26.48 & 4168 & 26.36 & 4.51\\
borg\_2140+0241        &  324.89 &  +2.69 & 3  & 0.076 & 1872 & 26.94 & 1406 & 26.32 & 1356 & 26.32 & 1156 & 26.50 & 1606 & 26.34 & 4.51\\  
borg\_2141-2310$^\ast$  &  325.15 & -23.17 & 3  & 0.042 & 1350 &       & 1556 &       & 1406 &       & 1256 &       & 1709 &       &    \\
borg\_2229-0945        &  337.19 &  -9.75 & 3  & 0.043 & 1479 & 26.83 & 1606 & 26.37 & 1506 & 26.34 & 1256 & 26.38 & 1759 & 26.25 & 4.49 
\enddata
\tablenotetext{}{NOTE: Column 1: Field name derived from the
  coordinates. Columns 2-3: $\alpha$ and $\delta$ coordinates (in
  degrees) as from the F140W exposure. Column 4: total number of {\it
    HST} orbits allocated. Column 5: Galactic extinction E(B-V) from
  \cite{schlafly2011}.  Columns 6-15: exposure time (in seconds) and
  $5\sigma$ limiting magnitude (in AB magnitudes) within a $r=
  0\farcs32$ aperture in each band. Column 16: effective area (in
  arcmin$^2$).}
\label{tab:field_characteristics}
\tablenotetext{$^\ast$}{Primary and parallel observations targeting
  M33 (borg\_0132+3035, borg\_0133+3043, borg\_0134+3034,
  borg\_0134+3041) or NGC7099 (borg\_2141-2310).}
\tablenotetext{$^\dagger$}{Images affected by scattered earthlight
  (see WFC3 Data Hand Book Section 6.10). We are currently building a
  model to successfully remove the effect from the data.}
\tablenotetext{$^\star$}{Guide star acquisition failure.}
\tablenotetext{$^\diamond$} {High Galactic extinction.}
\end{deluxetable*}
\clearpage

\begin{deluxetable*}{lcccccccccccccccc}[!h]
\tabletypesize{\footnotesize}
\tablecolumns{17}
\tablewidth{0pt} 
\tablecaption{\BORG\/ $z\sim9-10$ candidates}
\tablehead{\colhead{Obj ID} & \colhead{$\alpha$(J2000)} & \colhead{$\delta$(J2000)} & \colhead{\hh} & \colhead{$M_{AB}$} & \multicolumn{4}{c}{Colors}  & \multicolumn{5}{c}{$S/N$} & \colhead{$r_e$} & \colhead{Stellarity } & \colhead{Photo-$z$}\\
 & \colhead{(deg)} & \colhead{(deg)}& \colhead{(AB mag)} & \colhead{(AB mag)} &\colhead{\yy-\jj} & \colhead{\jj-\hh} &\colhead{\yy-\jh} & \colhead{\jh-\hh}& \colhead{F350LP} & \colhead{\yy} & \colhead{\jj} & \colhead{\jh} & \colhead{\hh} & & &}
\startdata
2134-0708\_774 & 323.5623 &  -7.1200 & 25.35$\pm$0.26 & -22.18$\pm$0.26 & $>0.37$       & 1.74$\pm$0.66 & $>1.53$       & 0.59$\pm$0.26  &  1.2  & 0.0 &  1.7 & 5.0  & 7.7  & 0.23 & 0.01 & 10.0\\
2140+0241\_37  & 324.8939 &  +2.6756 & 24.94$\pm$0.20 & -22.66$\pm$0.20  & -             & $>2.38$       & $>1.58$       & 0.79$\pm$0.27  & -1.2  & 0.1 & -0.3 & 4.7  & 8.28 & 0.37 & 0.01 & 10.5 \Bstrut\\
\hline                                                                 
0116+1425\_630 &  19.0347 & +14.4026 & 24.53$\pm$0.10 & -22.75$\pm$0.10  & 1.36$\pm$0.34 & 0.48$\pm$0.11 & 1.72$\pm$0.34 &  0.11$\pm$0.09 & -0.2 & 3.3 & 13.2 & 12.6 & 16.1 & 0.17 & 0.03 & 8.4\Tstrut \\
0956+2848\_85  & 149.1227 & +28.7920 & 26.41$\pm$0.19 & -20.91$\pm$0.19  & 1.70$\pm$2.60 & 0.45$\pm$0.27 & $>$1.65       &  0.05$\pm$0.21 & -0.5 & 0.4 &  4.9 &  7.7 &  7.2 & 0.08 & 0.07 & 8.7 \\
2229-0945\_548 & 337.1903 &  -9.7491 & 25.12$\pm$0.17 & -22.15$\pm$0.17  & 1.86$\pm$0.73 & 0.42$\pm$0.17 & 2.04$\pm$0.73 &  0.24$\pm$0.15 &  0.3 & 1.5 &  8.0 &  9.9 & 10.8 & 0.13 & 0.44 & 8.4
\enddata
\tablenotetext{}{NOTE: Coordinates and photometric properties of our
  $z\sim9$ and $z\sim10$ candidates. Columns 2-3: $\alpha$ and
  $\delta$ coordinates in degrees. Column 4: total magnitude in the
  \hh-band from {\tt SExtractor} MAG\_AUTO. Column 5: absolute
    magnitude. Columns 6-9: IR colors from {\tt SExtractor}
    MAG\_ISO. Columns 10-14: $S/N$ in each band.  Column 15: effective
    radius $r_e$ in arcseconds measured by {\tt SExtractor} and
    corrected for PSF. Column 16: stellarity index in \hh-band image
    from {\tt SExtractor} CLASS\_STAR. Column 17: photometric redshift
    obtained from BPZ.}
\label{tab:z9-10}
\end{deluxetable*}

\begin{deluxetable*}{lccccccccccccc}[!h]
\tabletypesize{\footnotesize}
\tablecolumns{14}
\tablewidth{0pt} 
\tablecaption{\BORG\/ $z\sim8$ candidates}
\tablehead{\colhead{Obj ID} & \colhead{$\alpha$(J2000)} & \colhead{$\delta$(J2000)} & \colhead{\hh} & \multicolumn{2}{c}{Colors}  & \multicolumn{5}{c}{$S/N$} & \colhead{$r_e$} & \colhead{Stellarity } & \colhead{Photo-$z$} \\
 & \colhead{(deg)} & \colhead{(deg)}& \colhead{(AB mag)} &\colhead{\yy-\jj} & \colhead{\jj-\hh}& \colhead{F350LP} & \colhead{\yy} & \colhead{\jj} & \colhead{\jh} & \colhead{\hh} & & & }
\startdata
0116+1425\_747 &  19.0372 & +14.4068 & 24.99$\pm$0.18 & 1.12$\pm$0.36 &  0.26$\pm$0.14 &  1.3 & 3.2 &  9.9 &  8.4 & 11.6 & 0.25 & 0.03 & 7.9 \\
0853+0310\_145 & 133.1855 &  +3.1467 & 25.26$\pm$0.14 & 0.68$\pm$0.15 & -0.07$\pm$0.12 & -1.0 & 8.7 & 14.6 & 16.5 & 12.3 & 0.08 & 0.25 & 7.6 \\
1103+2913\_1216& 165.6693 & +29.2273 & 26.12$\pm$0.19 & 0.53$\pm$0.23 & -0.11$\pm$0.19 &  0.9 & 5.5 &  8.7 &  8.4 &  7.3 & 0.17 & 0.01 & 7.3 \\
1152+3402\_912 & 177.9077 & +34.0397 & 25.20$\pm$0.23 & 0.75$\pm$0.23 &  0.17$\pm$0.15 &  1.0 & 5.3 &  9.5 & 11.6 & 10.6 & 0.18 & 0.03 & 7.6
\enddata
\tablenotetext{}{NOTE: Coordinates and photometric properties of our
  $z\sim8$ candidates. Columns 2-3: $\alpha$ and $\delta$ coordinates
  in degrees. Column 4: total magnitude in the \hh-band from {\tt
    SExtractor} MAG\_AUTO. Columns 5-6: IR colors from {\tt
    SExtractor} MAG\_ISO. Columns 7-11: $S/N$ in each band. Column 12:
  effective radius $r_e$ in arcseconds measured by {\tt SExtractor}
  and corrected for PSF. Column 13: stellarity index in the \hh-band
  image from {\tt SExtractor} CLASS\_STAR. Column 14: photometric
  redshift obtained from BPZ.}
\label{tab:z8}
\end{deluxetable*}
\end{turnpage}


\begin{deluxetable}{lcc}
\tablecolumns{3} 
\tablewidth{0pt} 
\tablecaption{\BORG\/ step-wise rest-frame UV LF at $z\sim8.7$ and $z\sim10$.}
\tablehead{\colhead{$z$} & \colhead{$M_{UV, AB}$} &
\colhead{$\phi$(10$^{-6}$ Mpc$^{-3}$mag$^{-1}$)} } \startdata
\multirow{4}{*}{8.7}    & -22.97 & $3.3^{+7.7}_{-2.7}$ \\
                        & -22.17 & $3.7^{+8.3}_{-3.1}$ \\
                        & -21.37 & $6.9^{+16.2}_{-5.6}$  \\
                        & -20.57 & $<110$\\
\hline
\multirow{4}{*}{10}  & -23.13 & $<4.7$\\
                     & -22.33 & $5.4^{+7.6}_{-3.5}$\\
                     & -21.53 & $<12$
\enddata 
\tablenotetext{}{NOTE: Upper limits are $1\sigma$.}
\label{tab:stepwiseLF}
\end{deluxetable}

\begin{turnpage}
\begin{deluxetable*}{lccccccccccccccc}[!h]
\tabletypesize{\footnotesize}
\tablecolumns{16}
\tablewidth{0pt} 
\tablecaption{\BORG\/ additional high-$z$ candidates}
\tablehead{\colhead{Obj ID} & \colhead{$\alpha$(J2000)} & \colhead{$\delta$(J2000)} & \colhead{\hh} & \multicolumn{4}{c}{Colors}  & \multicolumn{5}{c}{$S/N$} & \colhead{$r_e$} & \colhead{Stellarity } & \colhead{Photo-$z$} \\
 & \colhead{(deg)} & \colhead{(deg)}& \colhead{(AB mag)} &\colhead{\yy-\jj} & \colhead{\jj-\hh}& \colhead{\yy-\jh} & \colhead{\jh-\hh} &\colhead{F350LP} & \colhead{\yy} & \colhead{\jj} & \colhead{\jh} & \colhead{\hh} & & & }
\startdata
0751+2917\_1211& 117.7138 & +29.2929 & 26.97$\pm$0.35 & 0.53$\pm$0.20 & -0.66$\pm$0.23 & 0.43$\pm$0.20 & -0.56$\pm$0.23 & -0.1 & 6.3 & 10.3 & 10.0 &  5.2 & ps & 0.93 & 7.3 \\
1209+4543\_1696 & 182.3859 & 45.7250 & 24.96$\pm$0.40 & 1.47$\pm$0.51 &  0.31$\pm$0.13 & 2.15$\pm$0.51 & -0.37$\pm$0.10 &  0.1 & 2.2 & 10.4 & 16.2 & 15.2 & 0.28 & 0.03 & 8.6 \Bstrut \\
\hline
0853+0310\_912  & 133.1798 &  +3.1730 & 24.99$\pm$0.16 & 0.94$\pm$0.13 & 0.48$\pm$0.25 & 1.10$\pm$0.24 &  0.32$\pm$0.12 &  0.6 & 4.8 & 10.4 & 12.7 & 13.4 & 0.18 & 0.02 & 1.6 \Tstrut\\
0925+1360\_965  & 141.2946 & +14.0056 & 26.23$\pm$0.28 & 0.71$\pm$0.35 & 0.13$\pm$0.25 & 0.90$\pm$0.34 &  0.06$\pm$0.24 &  1.1 & 3.6 &  6.3 &  6.9 &  6.1 & 0.10 & 0.01 & 1.1 \\ 
1048+1518\_1140 & 161.9818 & +15.2961 & 26.40$\pm$0.20 & 1.04$\pm$0.63 & 0.68$\pm$0.29 & 1.52$\pm$0.60 &  0.20$\pm$0.21 & -0.1 & 1.9 &  4.5 &  7.3 &  7.4 & 0.08 & 0.19 & 1.9 \Bstrut \\
\hline
0119-3411\_22  &  19.7015 & -34.1830 & 25.06$\pm$0.15 & 0.84$\pm$0.31 &  0.38$\pm$0.17 & 1.10$\pm$0.28 &  0.12$\pm$0.13 &  0.4 & 4.0 &  7.5 & 12.7 & 11.8 & 0.14 & 0.04 & 1.5 \Tstrut
\enddata
\tablenotetext{}{NOTE: Coordinates and photometric properties of the
  additional dropout candidates that are discussed in Appendix A.
  Columns 2-3: $\alpha$ and $\delta$ coordinates in degrees. Column 4:
  total magnitude in the \hh-band from {\tt SExtractor}
  MAG\_AUTO. Columns 5-6: IR colors from {\tt SExtractor}
  MAG\_ISO. Columns 7-11: $S/N$ in each band. Column 12: effective
  radius $r_e$ in arcseconds measured by {\tt SExtractor} and
  corrected for PSF. Column 13: stellarity index in the \hh-band image
  from {\tt SExtractor} CLASS\_STAR. Column 14: photometric redshift
  obtained from BPZ.}
\label{tab:additional_sources}
\end{deluxetable*}
\end{turnpage}

\end{document}